%% file: main.tex
\documentclass[twocolumn]{aastex62}
\pdfoutput=1 
\usepackage{amsmath,amstext}
\usepackage[T1]{fontenc}
\usepackage[utf8]{inputenc}
\usepackage{apjfonts} 
\usepackage[figure,figure*]{hypcap}
\usepackage{amsmath,amssymb,amsfonts,graphicx}
\usepackage{todonotes}
\usepackage{color}
\usepackage{hyperref}
\usepackage{cleveref}
\usepackage{graphicx}
\usepackage{tabularx}
\usepackage{footnote}
\usepackage{subfigure}

\usepackage{todonotes}   

\newcommand{\RomanNumeralCaps}[1]
    {\MakeUppercase{\romannumeral #1}}

\shorttitle{PHOTOMETRY AND SPECTROSCOPY OF CONTACT BINARIES}
\shortauthors{Panchal \& Joshi}

\received{October 07, 2020}

\accepted{February 25, 2021}

\journalinfo{The Astronomical Journal}

\begin{document} 

\title{PHOTOMETRIC AND SPECTROSCOPIC ANALYSIS OF FOUR CONTACT BINARIES}

\author{Panchal, A.}
\affiliation{Aryabhatta Research Institute of observational sciencES (ARIES). Nainital, Uttrakhand, India.}
\affiliation{Department of Physics, DDU Gorakhpur University, Gorakhpur, India.}
\email{alaxender@aries.res.in}

\author{Joshi, Y. C.}
\affiliation{Aryabhatta Research Institute of observational sciencES (ARIES). Nainital, Uttrakhand, India.}

\begin{abstract}

   {}
   {We present the photometric and spectroscopic analysis of four W UMa binaries J015829.5+260333 (hereinafter as J0158), J030505.1+293443 (hereinafter as J0305), J102211.7+310022 (hereinafter as J1022) and KW Psc. The $VR_{c}I_{c}$ band photometric observations are carried out with the 1.3-m Devasthal Fast Optical Telescope (DFOT). For low resolution spectroscopy, we used 2-m Himalayan Chandra Telescope (HCT) as well as the archival data from 4-m LAMOST survey. The systems J0158 and J0305 show a period increase rate of $5.26(\pm1.72)\times10^{-7}~days~yr^{-1}$ and $1.78(\pm1.52)\times10^{-6}~days~yr^{-1}$, respectively. The period of J1022 is found to be decreasing with a rate of $4.22(\pm1.67)\times10^{-6}~days~yr^{-1}$. The period analysis of KW Psc displays no change in its period. PHOEBE package is used for the light curve modeling and basic parameters are evaluated with the help of GAIA parallax. The asymmetry of light curves is explained with the assumption of cool spots at specific positions on one of the components of the system. On the basis of temperatures, mass ratios, fill-out factors and periods, the system J1022  is identified as W-subtype systems while the others show some mixed properties. To probe the chromospheric activities in these W UMa binaries, their spectra are compared with the known inactive stars spectra. The comparison shows emission in $H_{\alpha}$, $H_{\beta}$ and Ca $\RomanNumeralCaps{2}$. To understand the evolutionary status of these systems, the components are plotted in mass-radius and mass-luminosity planes with other well characterized binary systems. The secondary components of all the systems are away from ZAMS which indicates that secondary is more evolved than the primary component.}
 

\end{abstract}

   
\section{Introduction} \label{sec:intro}
Eclipsing binaries (EBs) are the key sources to determine stellar parameters with high precision. One interesting class of EBs is contact binary stars (CBs). These are low temperature systems with components sharing a convective envelope. Due to the contact geometry, their temperatures are almost same and mostly they show equal size primary and secondary minima \citep{1941ApJ....93..133K, 1967PASP...79..395K, 1967AJ.....72S.309L, 1968ApJ...151.1123L}. The W UMa-type CBs (EWs) are particularly interesting as these are more abundant than other type of EBs \citep{1948HarMo...7..249S}. Secondly, the closeness of components of these systems allows us to directly perceive interaction between them and their atmosphere. Their orbital period is less than a day and both the components in EWs are located on or just above the main sequence with spectral type later than F \citep{1967PASP...79..395K, 1970PASJ...22..317O, 1972MNRAS.157..433M, 2005MNRAS.357..497B}. In most of the EWs deeper primary minima occurs when larger and more massive component passes in front of the smaller, less massive component. However, reverse can also occur in some cases. EWs are further divided into A and W-subtypes \citep{1970VA.....12..217B}. The A-subtype systems are earlier spectral type with higher mass and luminosity as compared to W-subtypes. In A-subtype systems, mass-ratio ($M_2/M_1$) is generally less than 0.5 and moderate or insignificant activity is observed. In W-subtype systems, less massive component is hotter and there is continuous change in the period with time \citep{1970VA.....12..217B, 1973AcA....23...79R}.

Many previous studies explain the origin of CBs from small period detached EBs (DEBs) \citep[e.g.,][]{1989A&A...220..128V, 2004MNRAS.355.1383L}. The loss of angular momentum (AML) due to magnetic braking is assumed to be leading formation mechanism for CBs \citep{2007ApJ...662..596L}. The ejection of mass due to magnetic activities can result in decrease in orbital or spin angular momentum, which can bring both components close to each other \citep{1966AnAp...29..331H, 1970PASJ...22..317O, 1982A&A...109...17V}. If AML continues even after contact phase, it can result in mass transfer between the components. Evolution of EWs depends upon AML, mass loss and mass transfer between the two components \citep{2012AcA....62..153S, 2013MNRAS.430.2029Y}. Analysing the LAMOST data for 7938 EWs, \cite{2017RAA....17...87Q} determined the parameters of CBs e.g., gravitational acceleration (log g), metallicity, temperature, radial velocity and found that about 80\% EWs have metallicity less than zero, which implies that EWs are old population systems. Many EWs are found to be magnetically active due to dynamo mechanism. The presence of magnetic field effects their evolution \citep{1967PASP...79..395K, 2008MNRAS.389.1722E}. Most of the EWs show asymmetrical light curves (LCs) i.e. difference in brightness at phases 0.25 and 0.75. This is generally explained by the presence of cool or hot spots on their surface. This effect is known as O'Connell effect \citep{1951PRCO....2...85O}. However, the amount of this asymmetry can change with course of time due to evolution and movement of spots on the stellar surface.

In this work, we present the multi-band photometric and low-resolution spectroscopic analysis of four EWs. These targets are chosen from Catalina Real Time Transient Survey (CRTS) which provides a catalog of $\sim47,000$ periodic variables \citep{2014yCat..22130009D}. Out of these variables $\sim31,000$ are classified as contact or ellipsoidal binaries. The J0158 ($\alpha_{2000}=01^{h}58^{m}29^{s}.5$, $\delta_{2000}=+26^{\circ}03^{\prime}33^{\prime\prime}$), J0305 ($\alpha_{2000}=03^{h}05^{m}05^{s}.1$, $\delta_{2000}=+29^{\circ}34^{\prime}43^{\prime\prime}$), J1022 ($\alpha_{2000}=10^{h}22^{m}11^{s}.7$, $\delta_{2000}=+31^{\circ}00^{\prime}22^{\prime\prime}$) and KW Psc ($\alpha_{2000}=22^{h}58^{m}31^{s}.7$, $\delta_{2000}=+05^{\circ}52^{\prime}23^{\prime\prime}$) are EWs, with approximate period of 0.227665, 0.246984, 0.2584680 and 0.234276 day, reported as in the CRTS Catalog. The list of targets and related information is given in Table~\ref{tar_info}.

\input{tab01.tex}

The paper is structured as follows: The information about photometric and spectroscopic observations is given in Section~\ref{Data}. The period estimation and period change is discussed in Section~\ref{orpe} which is followed by photometric analysis in Section~\ref{Ana}. The procedure used to determine physical parameters is described in Section~\ref{phy_para}. The spectroscopic analysis of these EWs is provided in Section~\ref{ch_ac}. The final results are discussed in Section~\ref{discu}.

\section{Observations}\label{Data}
\subsection{Photometry}\label{Photo}
The photometric observations of these targets have been acquired from the 1.3-m DFOT, Nainital employing a $2k\times2k$ CCD detector having a field of view of $\sim 18^{\arcmin}\times18^{\arcmin}$. As observations were carried out on different nights having varying moon illuminations, the exposure time varied across the frames. The total number of frames collected for J0158, J0305, J1022 and KW Psc were around 140, 200, 85 and 130, respectively in each band ($VR_{c}I_{c}$). Observing log for photometric observations is given in Table~\ref{log_phot}. 

\input{tab02.tex}
%

\input{tab03.tex}
%

\begin{figure*}[!ht]
\begin{center}
\includegraphics[width=17cm,height=7cm]{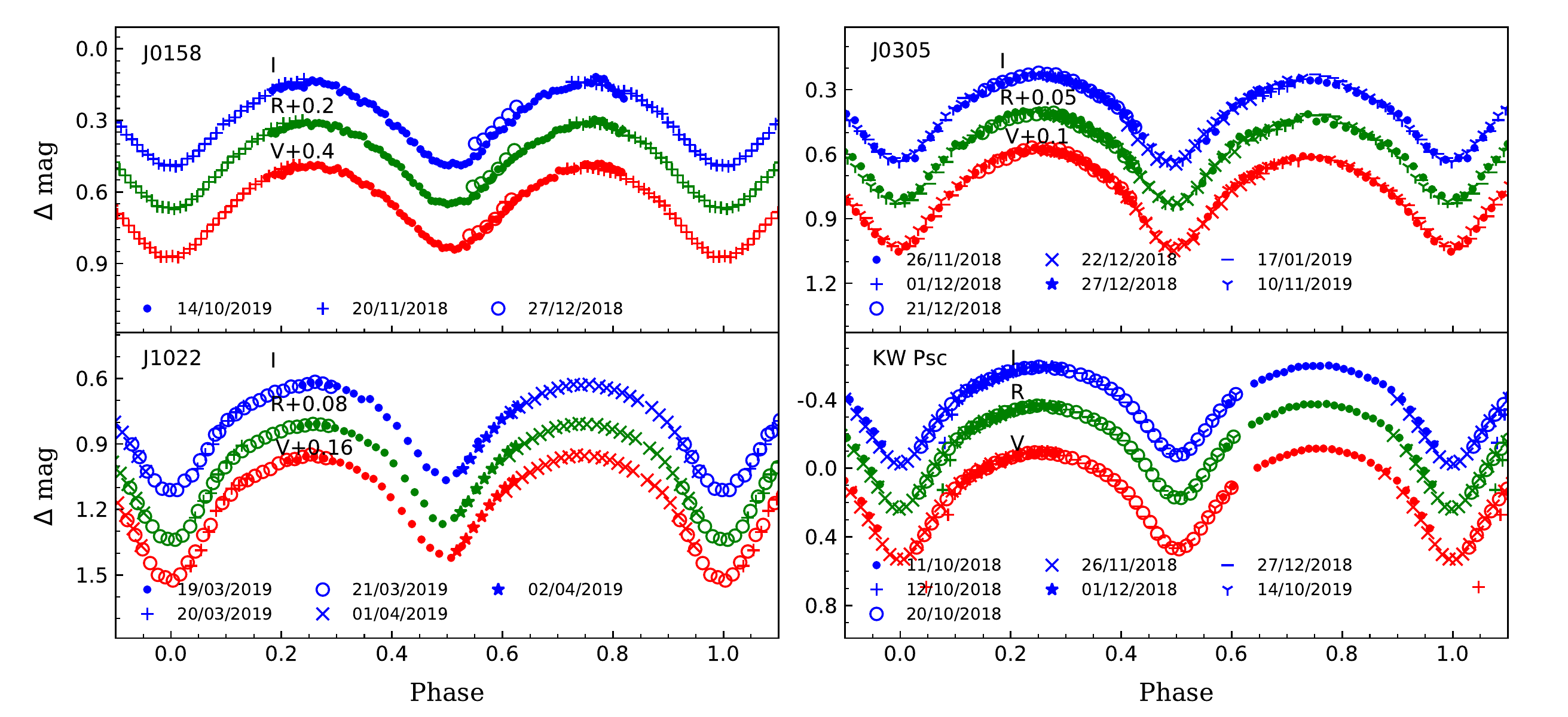}
\caption{The VRI band observed LCs of the sources. The different symbols show different date of observation.}
\label{lc_obs}
\end{center}
\end{figure*}

All the pre-processing steps like bias subtraction, flat fielding, cosmic ray removal were completed using IRAF routines. The instrumental magnitudes of target stars and comparison stars were computed by aperture photometry using DAOPHOT \citep{1992ASPC...25..297S}. Initially, five nearby field stars were selected having brightness similar to our targets for preparing differential LC. On the basis of differential LCs (Target star-Comparison star and Comparison star-Check star), most appropriate comparison stars and check stars were selected. For J0158, J0305, J1022 and KW Psc, we used TYC 1760-1359-1, TYC 1795-913-1, TYC 2510-242-1 and TYC 575-86-1 as comparison stars, respectively. The observed LCs in VRI bands are shown in Figure~\ref{lc_obs}.
\subsection{Spectroscopy}\label{Spec}
The Large sky Area Multi-Object Fibre Spectroscopic Telescope (LAMOST) is a 4-m aperture telescope with a field of view (FoV) of $5^{\circ} \times 5^{\circ}$. Such large FoV and a combination of 4000 fibers makes it a highly efficient tool for spectroscopy. It covers a spectral range of 370 nm to $\sim$ 900 nm with spectral resolution of 1 nm to $\sim$ 0.25 nm. The use of different gratings and camera positions can give a resolution (R) in range 1000-5000 \citep{2015RAA....15.1095L}. All the four targets were observed in the LAMOST survey. The data was downloaded for J0158 (1 spectra from DR3), J0305 (3 spectra from DR3 on different epochs), J1022 (3 spectra one each from DR1, DR2 and DR3) and KW Psc (1 spectra from DR1) from LAMOST website \footnote{http://dr5.lamost.org/}. The parameters mentioned in LAMOST database for these sources are given in Table~\ref{tar_lamost}. The spectral type was again estimated using the PyHammer, which uses empirical stellar spectra library with spectral types ranging from O5 to L3 and metallicity ranging from -2.0 dex to +1.0 dex. It covers a spectral range of 365 to 1020 nm \citep{2017ApJS..230...16K, 2020ascl.soft02011K}.

In addition to LAMOST spectroscopic data, Himalaya Faint Object Spectrograph Camera (HFOSC) on 2-m HCT was also used for observations. The observing log for these observations is given in Table~\ref{log_spec}. The combination of Gr7 and Gr8 grisms were used for observations. The Gr7 has a spectral range of 380-684 nm and a resolution of 1330. The Gr8 provides a wavelength range of 580-835 nm with resolution of 2190. For Gr7 spectra FeAr arc lamp and for Gr8 FeNe arc lamp were used for wavelength calibrations. For spectroscopic data reduction, IRAF package was used. Reduced calibrated spectra were normalized for further analysis. 

\input{tab04.tex}
\begin{figure*}[!ht]
\begin{center}
\includegraphics[width=15cm, height=7cm]{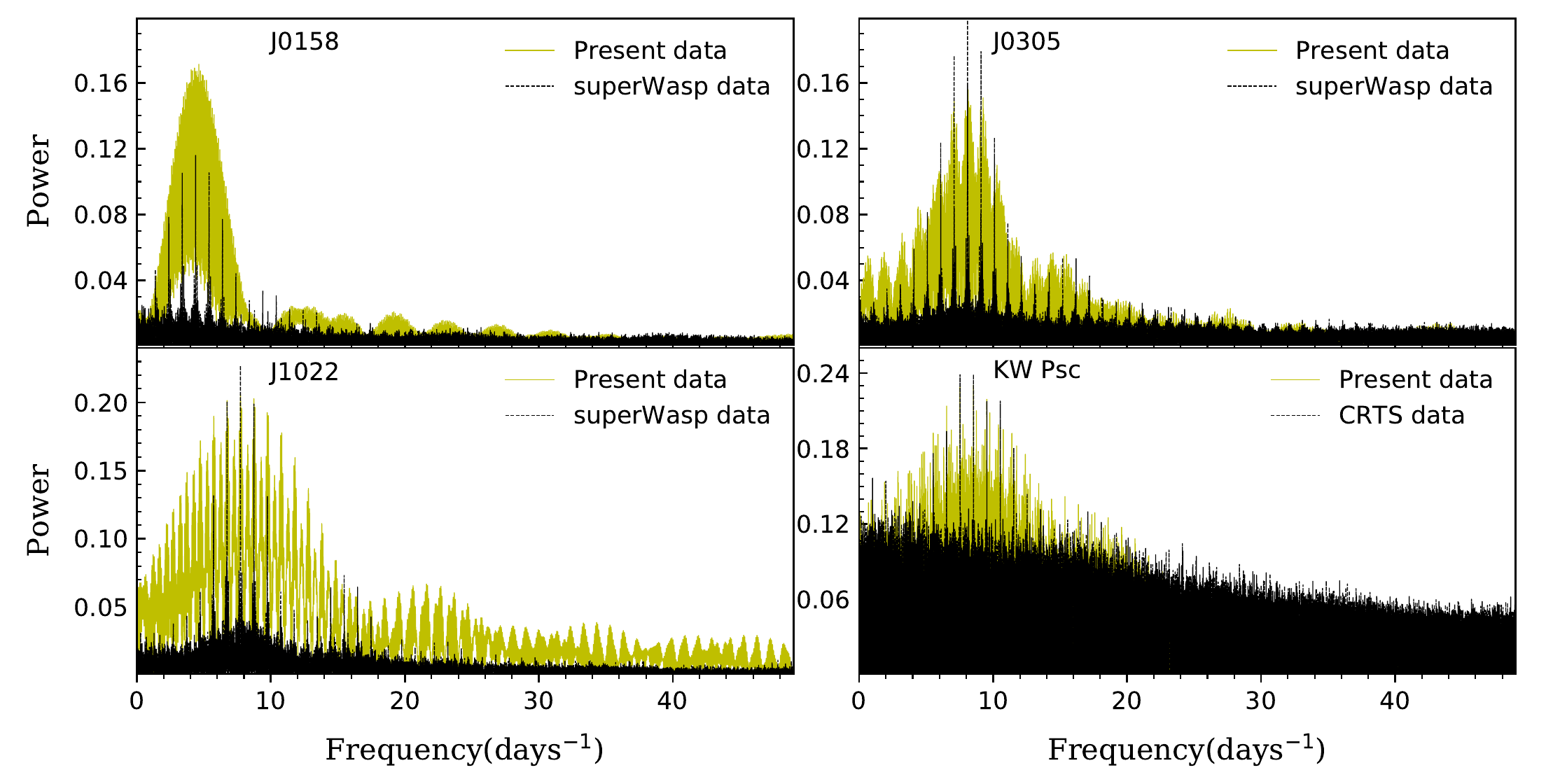}
\caption{Power spectra of four binary systems obtained using Period04. The power spectra obtained using SuperWASP data (for J0158, J0305 and J1022) and CRTS (for KW Psc) is over-plotted.}
\label{periodo}
\end{center}
\end{figure*}
%
\section{Orbital Period}\label{orpe}
The temporal variation in the orbital period of CBs provides useful information about mass transfer rate, presence of third body and other characteristics. Although \cite{2014yCat..22130009D} mentioned the approximate period of these systems, their periods were further determined with the present data using the Period04 software \citep{2004IAUS..224..786L}. Figure~\ref{periodo} shows the power spectra corresponding to all the four sources \textbf{using present data (green color) and archival data (black color)} The phase folded LC were plotted and visually analysed corresponding to these peaks. While for the systems J0305 and KW Psc, the best phase folded LCs were achieved corresponding to their highest peaks of power spectra, it was the nearby peaks close to the maximum peak in case of J0158 and J1022 which gave the best phase folded LCs. As the LC of CBs can be represented by twofold sine waves, the actual period of the system would be twice the period obtained from periodogram. The periods for J0158, J0305, J1022 and KW Psc are therefore found to be 0.447273 0.246982, 0.258484 and 0.234298 days, respectively.
\textbf{Since the spectra shown in Fig. 2 for each star are affected by strong side-lobes due to our short observing runs, we also obtained periodograms using SuperWASP data for J0158, J0305 and J1022 and CRTS data for KW Psc, which are over-plotted in Fig. 2. These periodograms show that periods obtained with the present data are very close to the periods determined with those of the archival data. We further ascertained our estimated periods through Period04 using the python periodogram based on the Lomb-Scargle method \citep{1976Ap&SS..39..447L, 1982ApJ...263..835S} and similar values were found}. While for later three systems, the newly estimated periods are close to earlier periods given by \cite{2014yCat..22130009D} but for J0158 newly estimated period is almost twice of that reported by \cite{2014yCat..22130009D}. The estimated period of J0158 is however a good match to those reported by \cite{2018yCat..22370028C} and \cite{2019yCat..51560241H}. The CRTS time series data used by \cite{2014yCat..22130009D} was reanalyzed and found that the power spectra of J0158 as represented by two sine waves indeed gave a period of 0.45 day.

The TOMs for primary or secondary eclipse were estimated with the help of Minima27 software \footnote{R.H. Nelson, www.variablestarssouth.org/resources/bob-nelson/} using the \cite{1956BAN....12..327K} method. To examine the period change, we searched for the multi-epoch photometric data or any available TOM information for these sources in the literature. Surveys like Catalina Sky Survey (CSS; \citealt{2014yCat..22130009D}), Wide Angle Search for Planets (SuperWASP; \citealt{2010A&A...520L..10B}), North Sky Variability Survey (NSVS; \citealt{2004AJ....127.2436W}), All Sky Automated Survey for SuperNovae (ASAS-SN; \citealt{2018MNRAS.477.3145J}) and others provide a good database of photometric data. The three sources (J0158, J0305 and J1022) were observed in these surveys but with the poor cadence. For systems J0305 and J1022, we were able to find half or complete phase of LCs on different days in SuperWASP data as their period is around 0.25 days. But for J0158, we could get only half LCs on different days as its period is $\sim10$ hrs. We also constructed the LCs for these three sources from CSS multi-epoch data as CSS time resolution was less than the SuperWASP. The system KW Psc was not observed in any of the above surveys although we found 19 TOMs available for this system on O-C gateway \footnote{http://var2.astro.cz/ocgate/}. In the following sub-sections, we individually analyze the four sources using their accumulated data.

\subsection{J0158}\label{J0158_per_stu}
\begin{figure*}[!ht]
\label{oc_0158_0305}
\begin{center}
\subfigure{\includegraphics[width=8cm,height=7cm]{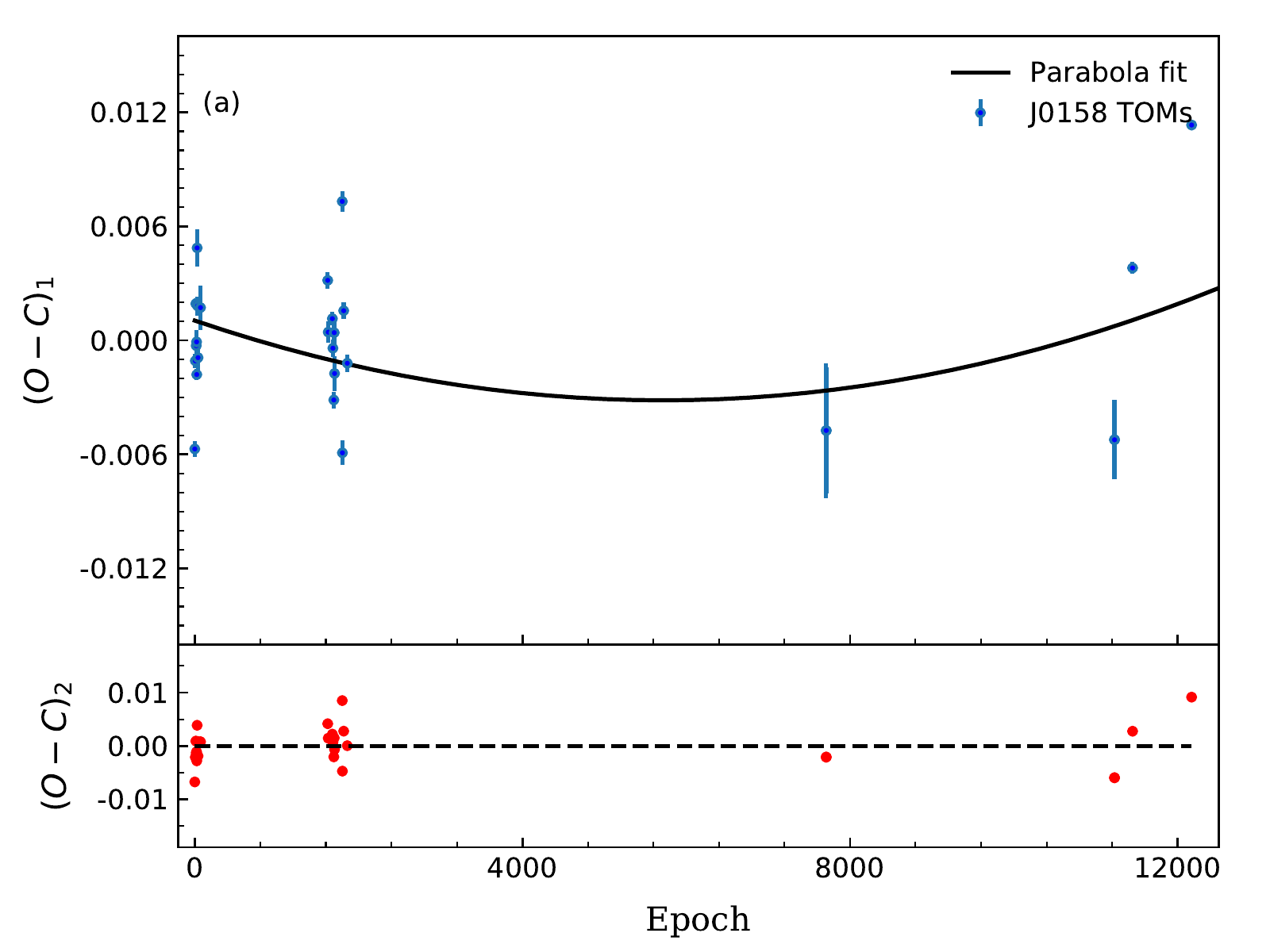}}
\subfigure{\includegraphics[width=8cm,height=7cm]{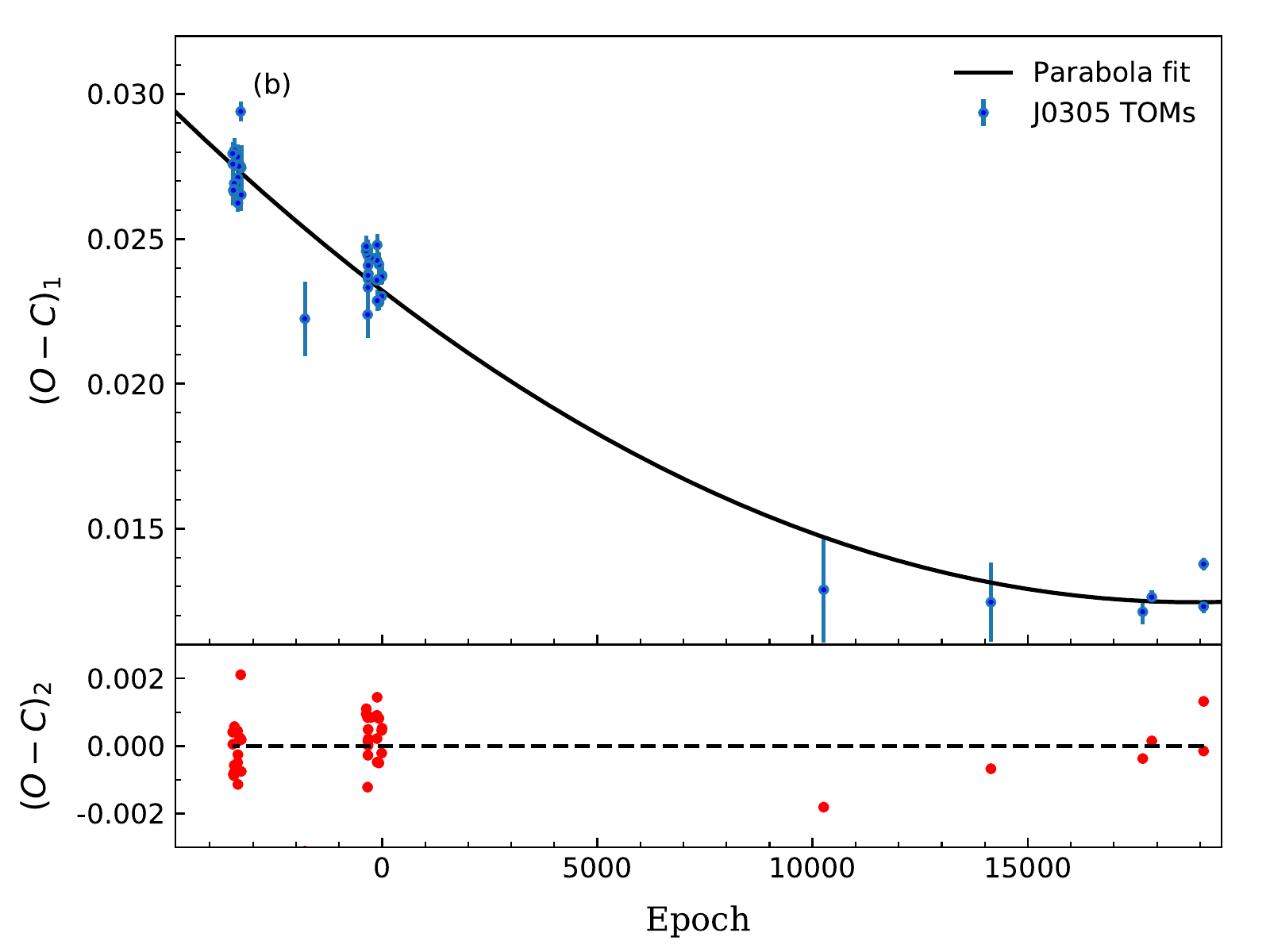}}
\caption{O-C diagrams for (a) J0158 and (b) J0305 with quadratic regression. The lower panels show the residuals of the fit.}
\end{center}
\end{figure*}
For J0158, a total of 27 TOMs (21 TOMs from SuperWASP data, 4 from ASAS data and 2 from our data) were determined, as given in sample Table~\ref{OC_info}. The updated linear ephemeris is estimated as:
\begin{equation}
\label{li_58}
HJD_{o}=2453229.6847(\pm0.0012)+0.4553331(\pm0.0000002)\times E
\end{equation} 
Here, $HJD_{o}$ represents TOM corresponding to primary minima and E is the number of epoch. The quadratic fit to the $(O-C)_{1}$ is shown in Figure~\ref{oc_0158_0305} (a). The $(O-C)_{1}$ shows an upward parabolic variation as shown in Figure~\ref{oc_0158_0305} (a) which can be represented by the following equation:
\begin{equation}
\label{qu_58_oc}
\begin{aligned}
(O-C)_{1} &=0.00104(\pm0.00118)-1.46736(\pm0.90607)\times 10^{-6} \times E \\
&+ 1.28313(\pm0.77219) \times 10^{-10} \times E^{2}
\end{aligned}
\end{equation}
This trend shown by J0158 suggests a continuous increase in its period. The modified quadratic ephemeris can therefore be expressed as:
\begin{equation}
\label{qu_58}
\begin{aligned}
HJD_{o} &=2453229.6859(\pm0.0015)+0.455329(\pm0.000001) \times E \\
&+ 3.28(\pm1.07) \times 10^{-10} \times E^{2}
\end{aligned}
\end{equation}
On the basis of above equation, the rate of period increase was estimated as  $5.26(\pm1.72)\times10^{-7}~days~yr^{-1}$ for the system J0158. The change in orbital period for contact binaries is normally due to the mass-transfer or mass-loss from one component to the other which can be calculated from the relation given by \cite{1958BAN....14..131K}.
\begin{equation}
 \label{matr}
 \frac{1}{M_{1}}\dfrac{dM_{1}}{dt}=\dfrac{q}{3(1-q)}\dfrac{1}{P}\dfrac{dp}{dt}
\end{equation}
Here, $q$ is the mass ratio defined by $M_2/M_1$. The above equation suggests that for a system with increasing period, the $dM_{1}$ will be negative if $q>1$ and positive if $q<1$. If the period of system is decreasing then $q>1$ will result in positive $dM_{1}$ and vice-versa. The negative $dM_{1}$ corresponds to mass transfer from primary component to the secondary component. The positive period change rate for J0158 along with $q < 1$ as determined in Section~\ref{Ana} suggests that the mass transfer is taking place from the secondary to primary component. Using above equation, the mass transfer rate for J0158 was estimated as $9.866\times10^{-7}~M_{\odot}~yr^{-1}$. The $M_{1}$ used in above equation in determined in Section~\ref{phy_para}.
\input{tab05.tex}
\begin{figure*}[!ht]
\begin{center}
\label{oc_1022_Psc}
\subfigure{\includegraphics[width=8cm, height=7cm]{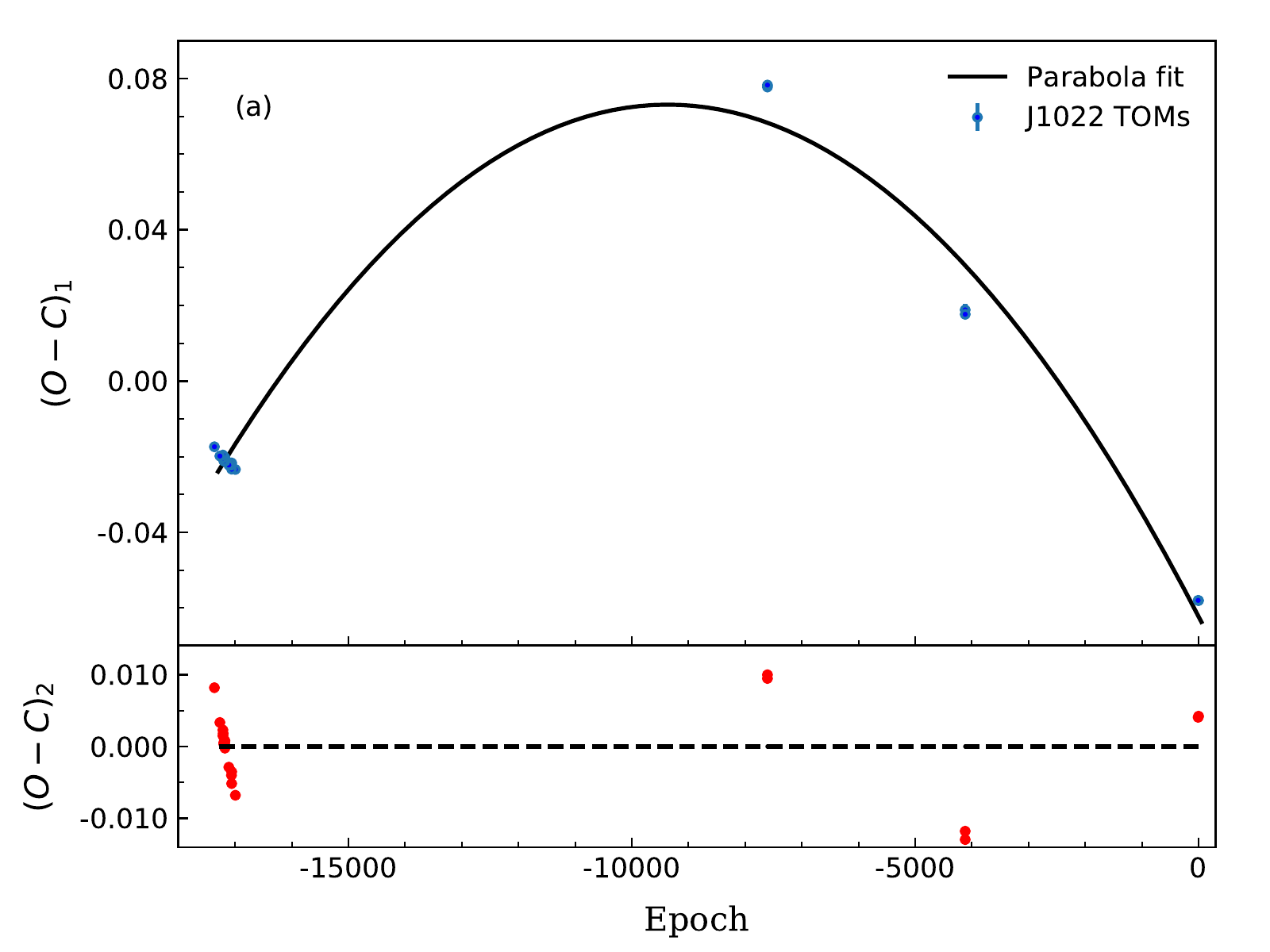}}
\subfigure{\includegraphics[width=8cm, height=7cm]{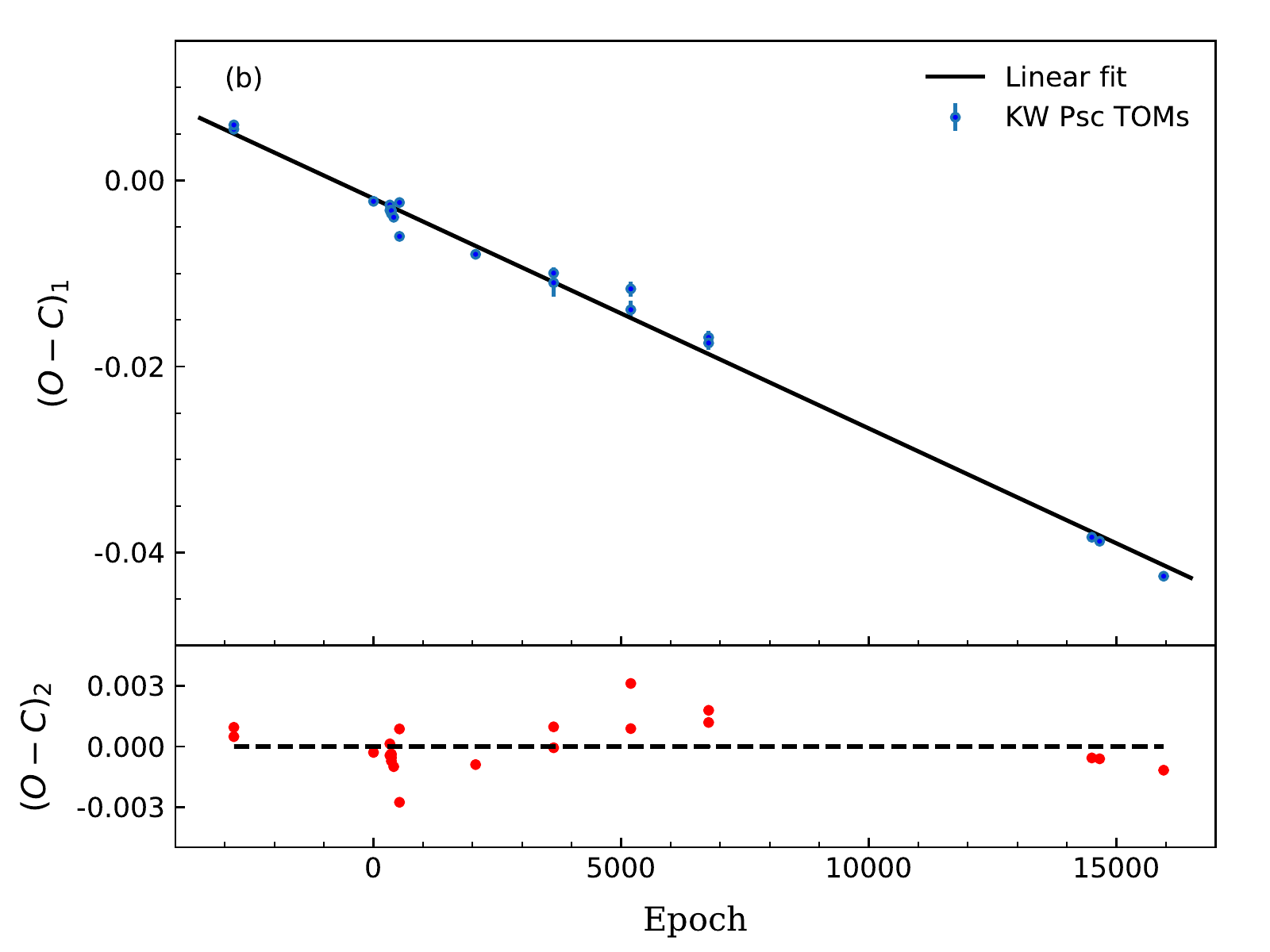}}
\caption{O-C diagrams for (a) J1022 and (b) KW Psc with quadratic  and linear regression, respectively. The lower panels show the residuals.}
\end{center}
\end{figure*}
\subsection{J0305}\label{J0305_per_stu}
For J0305, we were able to estimate 41 TOMs which comprises 32 from SuperWASP, 1 from CSS, 4 from ASAS and 4 from our data. The corresponding (O-C) diagram with a quadratic fit is shown in Figure~\ref{oc_0158_0305} (b). Like J0158, this system also shows an upward parabolic trend. The updated linear ephemeris for J0305 is given by:
\begin{equation}
\label{li_05}
HJD_{o}=2454085.436(\pm0.017)+0.246983(\pm0.000002)\times E
\end{equation} 
The modified quadratic ephemeris for J0305 was determined as:
\begin{equation}
\label{qu_05}
\begin{aligned}
HJD_{o} &=2454085.418(\pm0.023)+0.246974(\pm0.000008) \times E \\
&+ 6.02(\pm5.14) \times 10^{-10} \times E^{2}
\end{aligned}
\end{equation}
Similarly, the second order polynomial fitted to $(O-C)_{1}$ as shown in Figure~\ref{oc_0158_0305} (b) is as follows:
\begin{equation}
\label{qu_05_oc}
\begin{aligned}
(O-C)_{1} &=0.02323(\pm0.00020)-1.14030(\pm0.07074)\times 10^{-6} \times E \\
&+ 3.01937(\pm0.45643) \times 10^{-11} \times E^{2}
\end{aligned}
\end{equation}
Using the quadratic ephemeris equation, we found that the rate of period change for J0305 is $1.78(\pm1.52)\times10^{-6}~days~yr^{-1}$. We used Equation~\ref{matr} to determine the mass transfer rate in J0305 which is found to be $1.001\times10^{-6}~M_{\odot}~yr^{-1}$. The increasing period and $q<1$ for J0305 indicates the mass transfer from secondary to primary component.
\subsection{J1022}\label{J1022_per_stu}
For J1022, we estimated 22 TOMs that includes 16 from SuperWASP, 4 from ASAS and 2 from our data. The (O-C) diagram for J0122 is shown in Figure~\ref{oc_1022_Psc} (a). The quadratic fit shows downward parabolic variation. This means that period of J1022 is decreasing with time. The updated linear ephemeris for J1022 is found to be
\begin{equation}
\label{li_22}
HJD_{o}=2458564.233(\pm0.033)+0.258484(\pm0.000002)\times E
\end{equation} 
From the (O-C) diagram, the updated quadratic ephemeris for J1022 was found as follows:
\begin{equation}
\label{qu_22}
\begin{aligned}
HJD_{o} &=2458564.170(\pm0.039)+0.258455(\pm0.000011) \times E \\
&-1.494(\pm0.593) \times 10^{-9} \times E^{2}
\end{aligned}
\end{equation}
The rate of decrease in period, according to the above equation, is found to be $4.22(\pm1.67)\times10^{-6}$ days/year. The decrease in period can be attributed to AML via magnetic braking, gravitational wave radiation (GWR) or mass loss/transfer. We also estimated period decay rate due to GWR and magnetic braking which are estimated using equations given by \cite{1962ApJ...136..312K} and \cite{1988ASIC..241..345G}, respectively. The period decrease rate due to GWR corresponds to $2.486\times 10^{-16}$ days/year which is very small in comparison to the observed rate. The period decrease due to magnetic braking is found to be $7.005\times 10^{-8}$ days/year which is $\sim2\%$ of the observed period decay rate. Therefore most plausible mechanism behind the observed period change could be the mass transfer between the two components. We obtained a mass transfer rate of $2.467\times10^{-6}M_{\odot}$ per year from secondary to the primary which can explain the period change in the system J0122.
\subsection{KW Psc}\label{KWPsc_per_stu}
As mentioned earlier, KW Psc was not observed in any of the surveys except ASAS. From previous studies \citep{2010IBVS.5922....1G, 2010IBVS.5920....1D, 2011IBVS.5960....1D, 2012IBVS.6011....1D, 2013IBVS.6042....1D}, 23 TOMs were collected for KW Psc. From these values, the updated linear ephemeris is estimated as:
\begin{equation}
\label{li_Psc}
HJD_{o}=2455014.845(\pm0.008)+0.234278(\pm0.000001)\times E
\end{equation} 
The (O-C) diagram for KW Psc is shown in Figure~\ref{oc_1022_Psc} (b) and the residual plot is shown in the lower panel of Figure~\ref{oc_1022_Psc} (b). The $(O-C)_{1}$ for KW Psc can be written as:
\begin{equation}
\label{qu_Psc}
\begin{aligned}
(O-C)_{1} &=-0.00195(\pm0.00030)-2.47017(\pm0.04911) \times E \\
\end{aligned}
\end{equation}
The O-C diagram is a straight line with negative slope which means that its period is almost constant over a period of atleast 12 years during 2007 to 2019.
\section{Photometric Analysis}\label{Ana}
For photometric analysis of LCs in different bands, we used PHOEBE-1.0 (PHysics Of Eclipsing BinariEs) package \citep{2005ApJ...628..426P}. It is an open source modeling program based on Wilson-Devinney code \citep{1971ApJ...166..605W} for computing theoretical photometric and radial velocity curves in the binary systems. It can work with two different minimization algorithms namely differential corrections and Nelder $\&$ Mead's Simplex. In present analysis, differential corrections minimization algorithm was used. As present systems are reported to be EWs, hence, the over-contact binary not in thermal contact mode was used during their photometric analysis.
\subsection{Effective Temperature}\label{teff}
Although these sources were selected from \cite{2014yCat..22130009D} but these were also observed in other surveys like SuperWASP, ASAS, KELT, 2MASS, NSVS etc. as discussed in Section~\ref{periodo}. Their magnitudes in $B$, $V$, $J$, $H$ and $K$ bands were collected from available archival catalogs. We calculated the effective temperature ($T_{eff}$) using the (J-H)-$T_{eff}$ relation from \cite{2007MNRAS.380.1230C} as given below.
\begin{equation}
T_{eff}=-4369.5(J-H)+7188.2
\end{equation}
Here, $(J-H)$ color index is taken from 2MASS and $T_{eff}$ represents effective temperature of the star. For J0158, J0305, J1022 and KW Psc, the $T_{eff}$ is determined as 6140 ($\pm$105), 4829 ($\pm$105), 5440 ($\pm$118) and 5047 ($\pm$138) K, respectively using the above equation. The $T_{eff}$ is also calculated with $(B-V)_{o}-T_{eff}$ relations given by \cite{1994ApJ...434..277W} and \cite{2010AJ....140.1158T}. The $T_{eff}$ obtained from different equations as well as those provided in the LAMOST survey are listed in Table~\ref{all_temp}. It can be seen from the table that the $T_{eff}$ obtained using different methods are almost similar for all sources except J0305. Finally, we calculated the average temperature and used it as $T_{eff}$ for the primary component during LC model fitting.
\input{tab06.tex}
%
\begin{figure*}[!ht]
\begin{center}
\includegraphics[height=7cm,width=14cm]{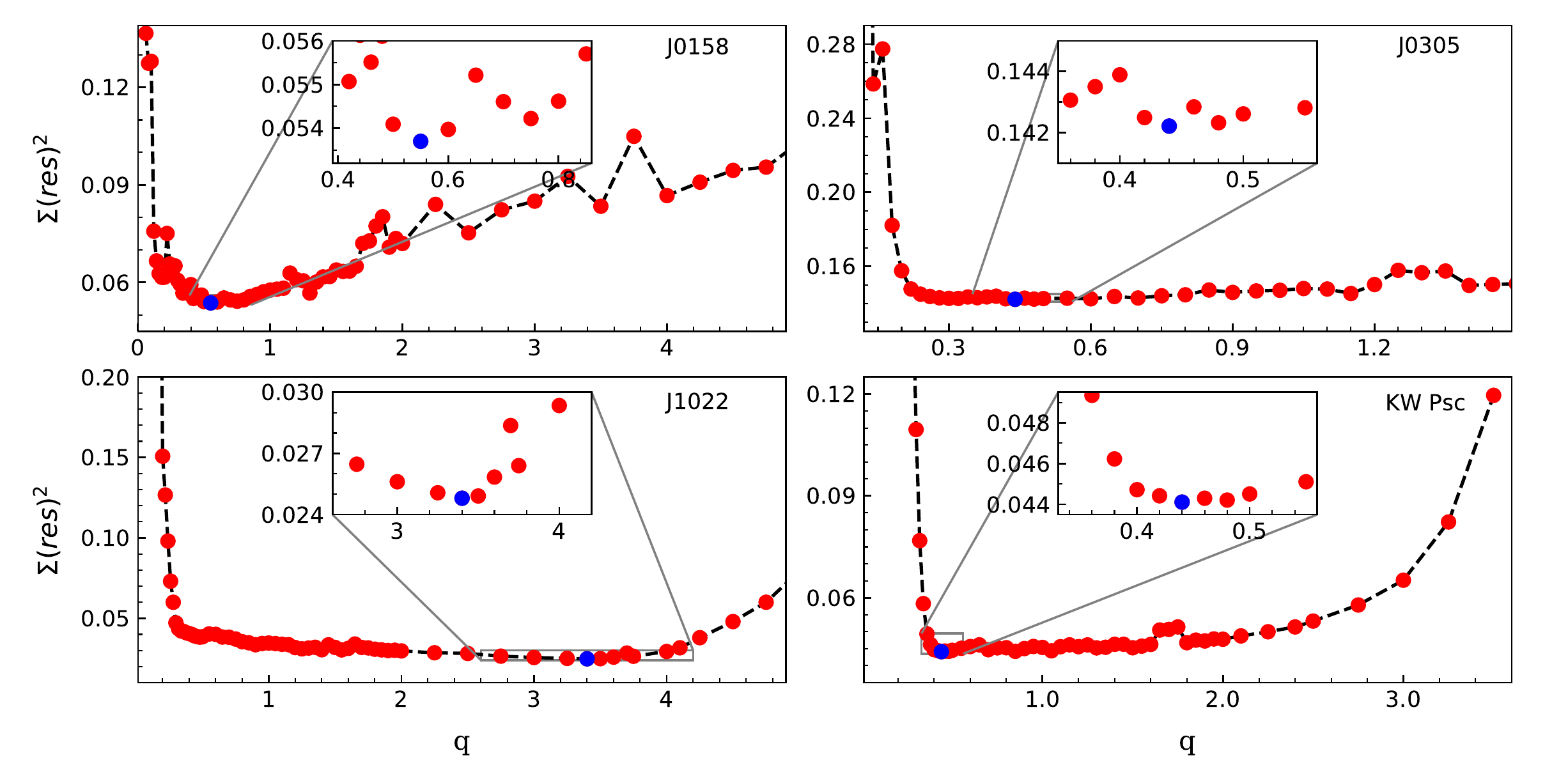}
\caption{The estimation of q-parameter for the EWs marked at the top right corner of each panel.}
\label{qsearch}
\end{center}
\end{figure*}
\subsection{q-search and Modeling}\label{q_param}
The accurate determination of mass-ratio requires multi-epoch radial velocity (RV) information for each component. Due to absence of RV data, q-search technique was used for the estimation of $q$ parameter from the photometric data \citep[e.g.,][]{2016RAA....16...63J, 2017RAA....17..115J}. In this process, we fixed the gravity darkening coefficients as $g_{1}=g_{2}=0.32$ and bolometric albedo as $A_{1}=A_{2}=0.5$ assuming these EWs having convective envelopes. The limb darkening coefficients were estimated automatically by the program using tables from \cite{1993AJ....106.2096V} with square root limb-darkening laws. We set the $T_{eff}$ of primary as determined in Section~\ref{teff}. Then, we varied the $q$ parameter from 0.1 to higher values in steps of 0.02-0.05 and ran the PHOEBE program corresponding to each value of q. In this process, other parameters like secondary $T_{eff}$, primary component surface potential ($\Omega_{1}$), primary component luminosity ($L_{1}$) and inclination ($i$) are set as free parameters. The sum of squared residuals ($\Sigma res^2$) obtained corresponding to best fit is plotted verses corresponding $q$. The Figure~\ref{qsearch} shows that the solution is converged at some specific value of $q$ corresponding to minimum $\Sigma res^2$ for each system. The $q$ is estimated as 0.55($\pm$1), 0.44($\pm$2), 3.40($\pm$1) and 0.44($\pm$1) for J0158, J0305, J1022 and KW Psc, respectively.  

\begin{figure*}[!ht]
\begin{center}
\label{uniq}
\subfigure{\includegraphics[width=6cm, height=4cm]{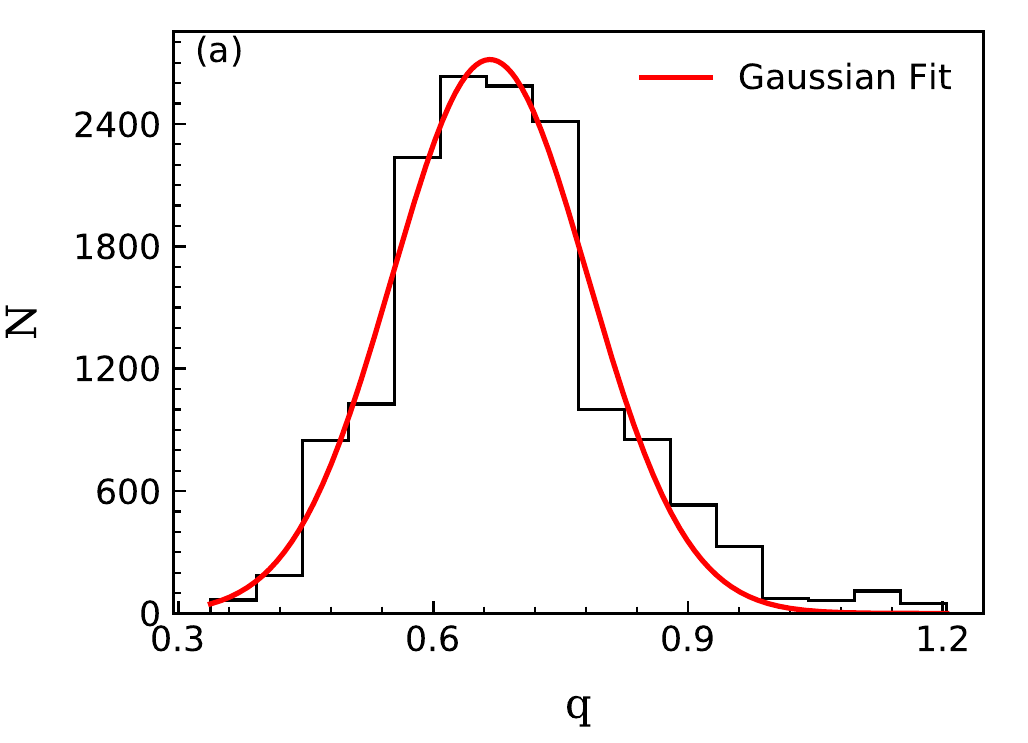}}
\subfigure{\includegraphics[width=6cm, height=4cm]{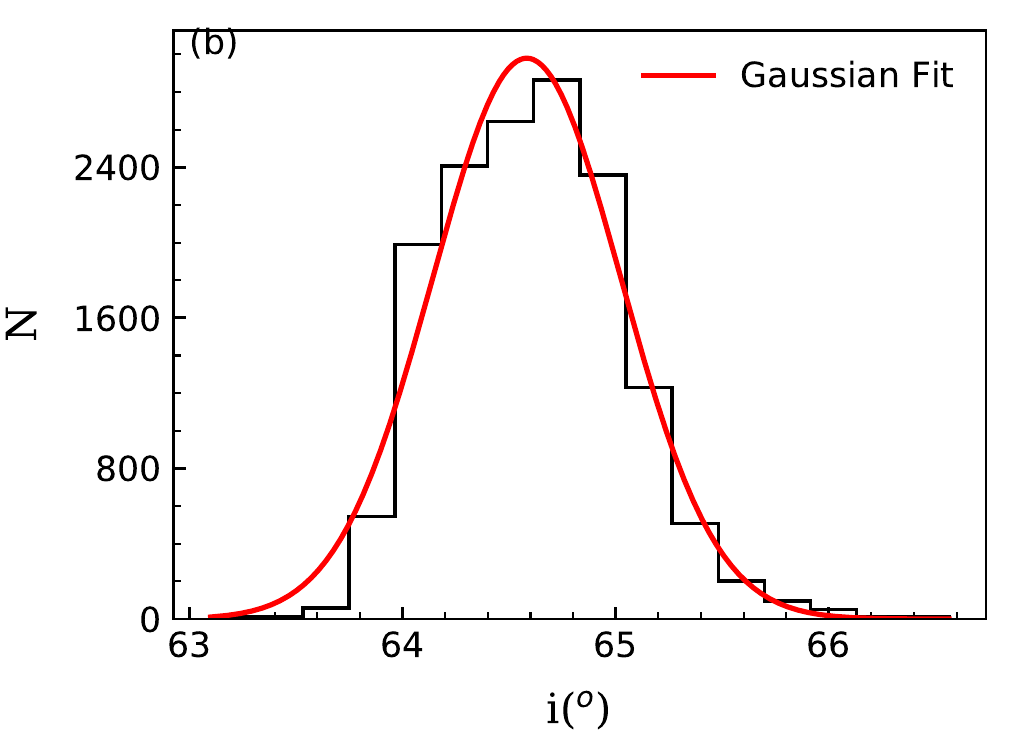}}\vspace{-0.3cm}
\subfigure{\includegraphics[width=6cm, height=4cm]{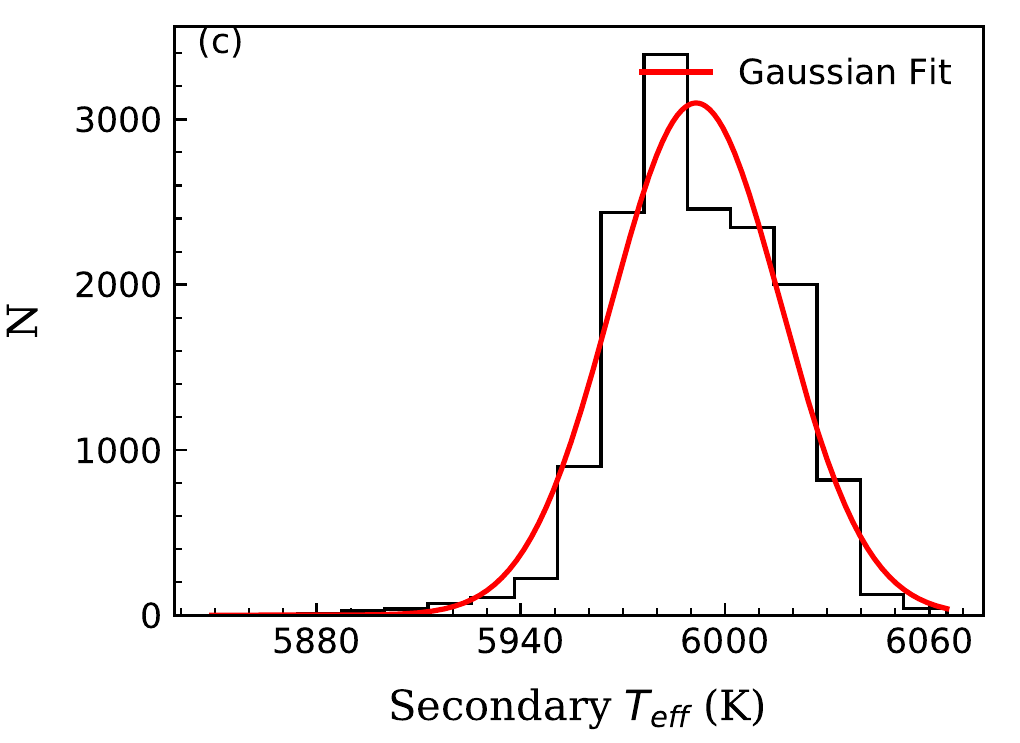}}
\subfigure{\includegraphics[width=6cm, height=4cm]{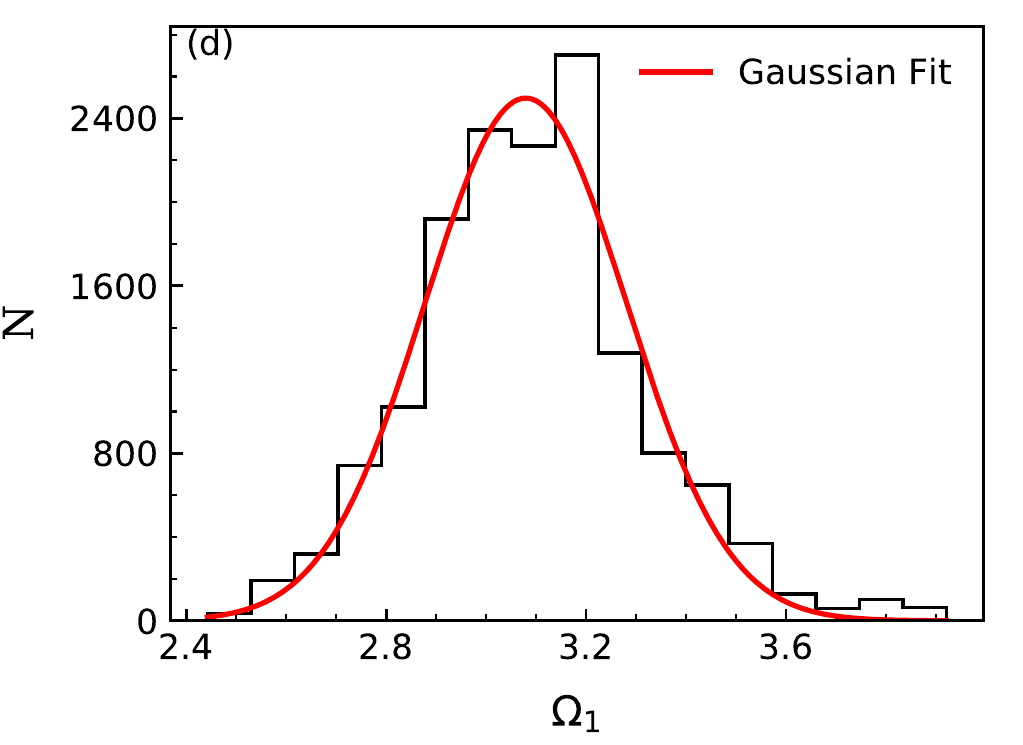}}
\caption{Hitograms obtained for four parameters using heuristic scanning and parameter kicking.}
\end{center}
\end{figure*}

\begin{figure*}[!ht]
\begin{center}
\label{scan}
\subfigure{\includegraphics[width=7cm, height=4cm]{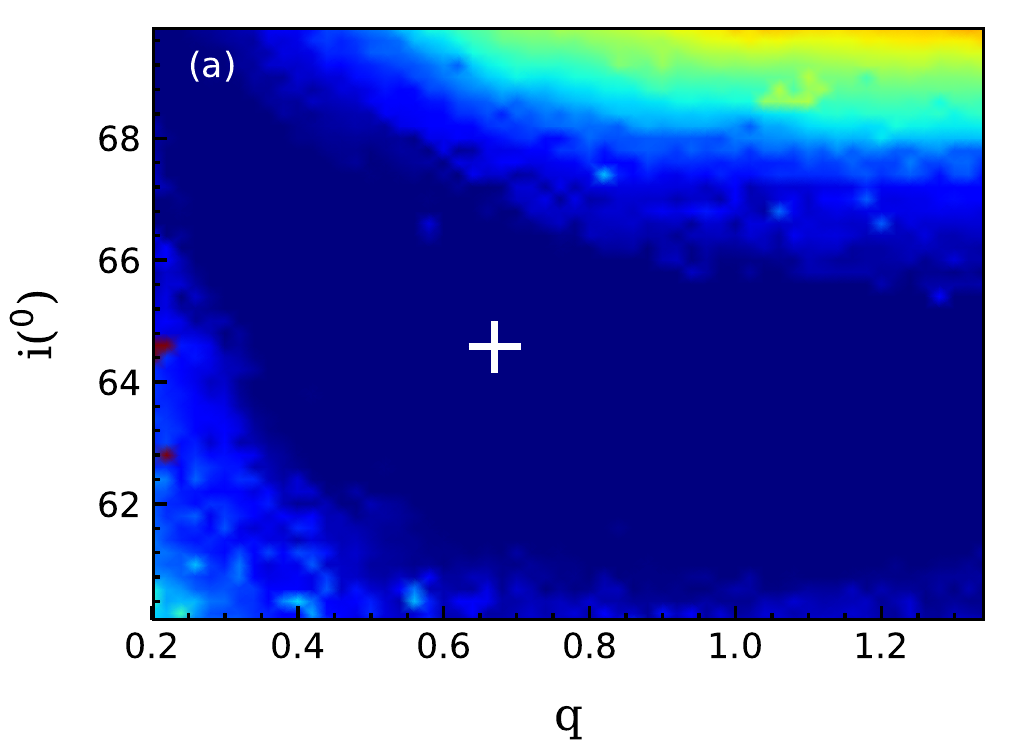}}
\subfigure{\includegraphics[width=7cm, height=4cm]{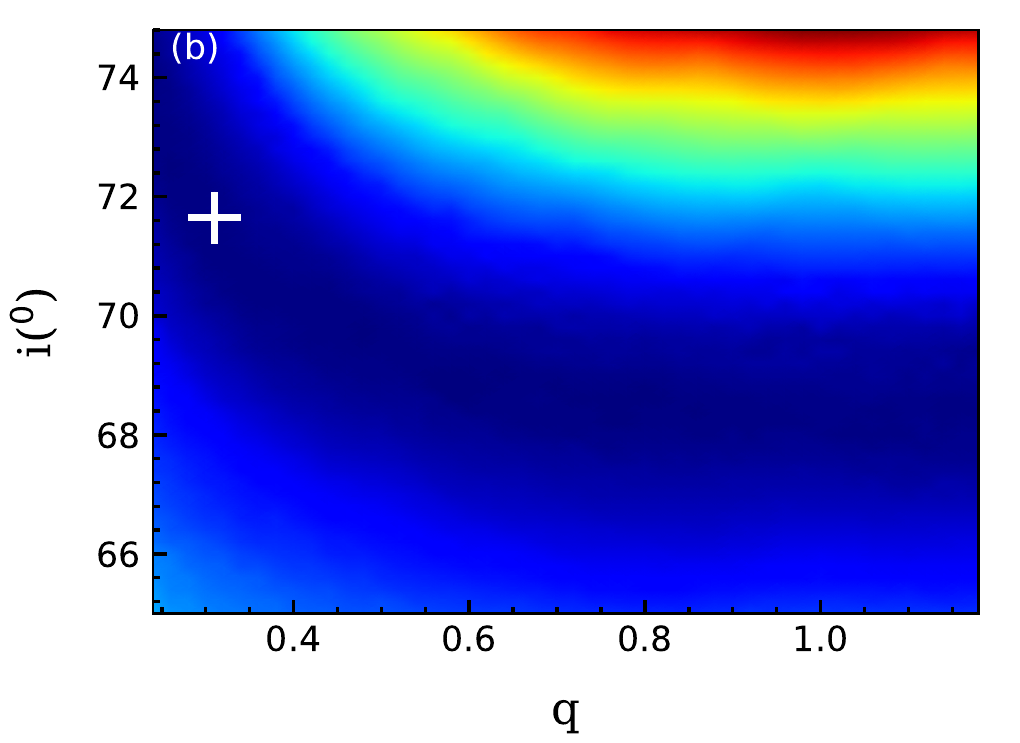}}\vspace{-0.3cm}
\subfigure{\includegraphics[width=7cm, height=4cm]{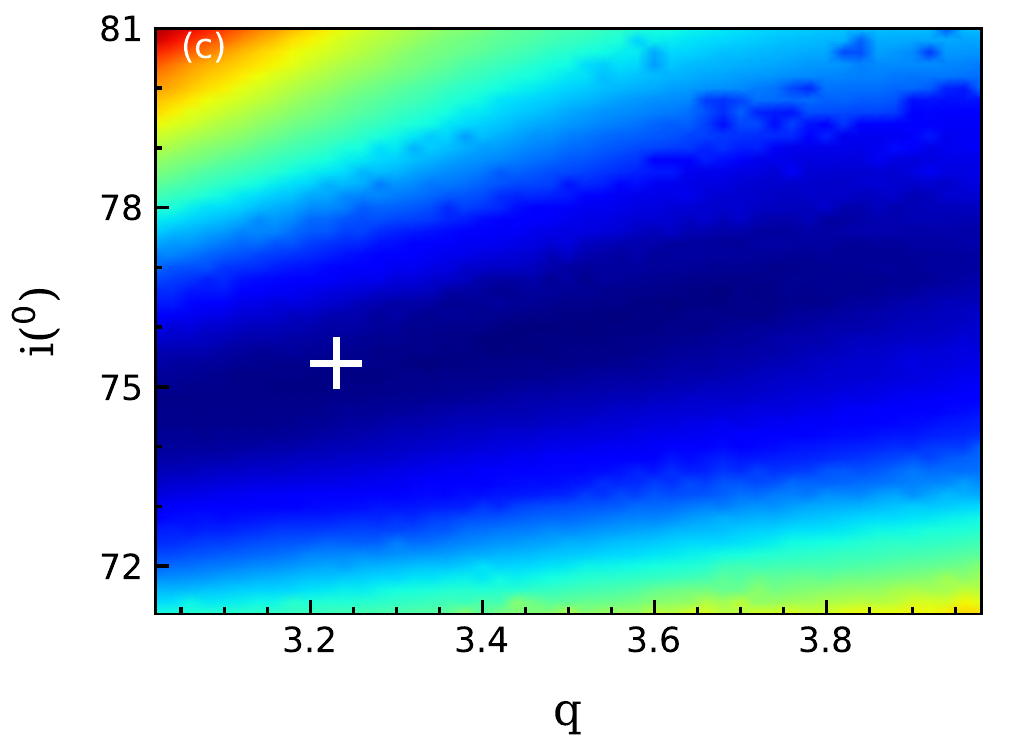}}
\subfigure{\includegraphics[width=7cm, height=4cm]{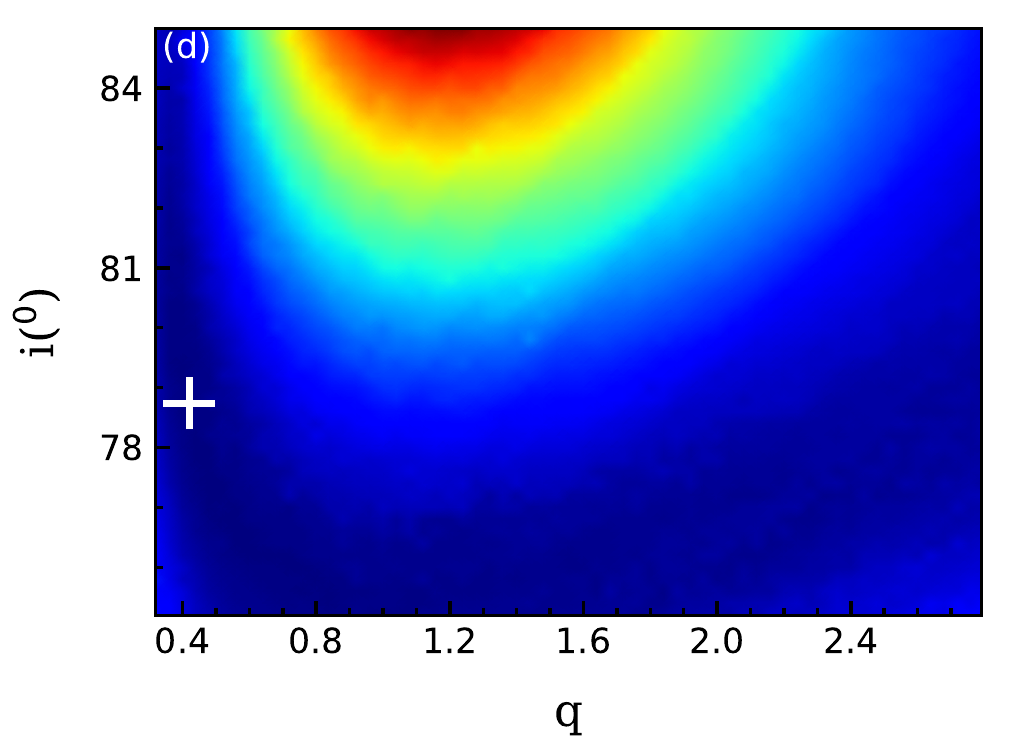}}
\caption{The q-i parameter space mapping for Figure~\ref{scan} (a). J0158, Figure~\ref{scan} (b). J0305, Figure~\ref{scan} (c). J1022 and Figure~\ref{scan} (d). KW Psc. The + sign represents the final model q-i. The low chi-square regions are represented by blue color and red color represents high chi-square regions.}
\end{center}
\end{figure*}

The best q and corresponding parameters obtained in q-search are initial estimates. The Figure~\ref{qsearch} shows that q-search has given a wide range of equiprobable q for J0305, J1022 and KW Psc. The final parameters, associated errors and uniqueness of the these solutions were explored with the help of PHOEBE scripter. The scripter was run for 15000 iterations with differential correction minimizations. All the parameters e.g. i, q, secondary $T_{eff}$, $\Omega_{1}$, $\Omega_{2}$ and L were set free with initial values obtained during q-search process. After every 50 iterations a kick of $\pm5\%$ was introduced to all parameters. The fit converged to minimum after these kicks after 5-10 iterations. The output was saved after each iteration. The final values were determined by guassian fitting to the histograms of these iteration results. The Figure~\ref{uniq} shows some of the gaussian fitted histograms. For all four systems, the estimated parameters were almost similar to the best fit parameters obtained during q-search process. The final $q$ for J0158, J0305, J1022 and J2258 are found to be 0.67, 0.31, 3.23, 0.42, respectively, after the gaussian fitting.
%
%
The heuristic scanning was used for checking the stability of adopted solution in the nearby parameter space. Almost 50-60 values of $q$ and $i$ within $\pm5\%$ of the above obtained values were used to generate a grid of $\sim$2500-3000 perturbed models. Figure~\ref{scan} represents the color map of $q$ versus $i$ in these models. It is a 2-D histogram representing the variation of chi squares in the q-i parameter hyperspace obtained by heuristic scanning. The blue end of the color scale represents the minimum chi square. The "+" signs indicates the position of final adopted model in q-i space. It can be seen that determined models are in the bluer regions which corresponds to a better fit model.
%
%
\begin{figure*}[!ht]
\begin{center}
\includegraphics[width=16.5cm, height=7cm]{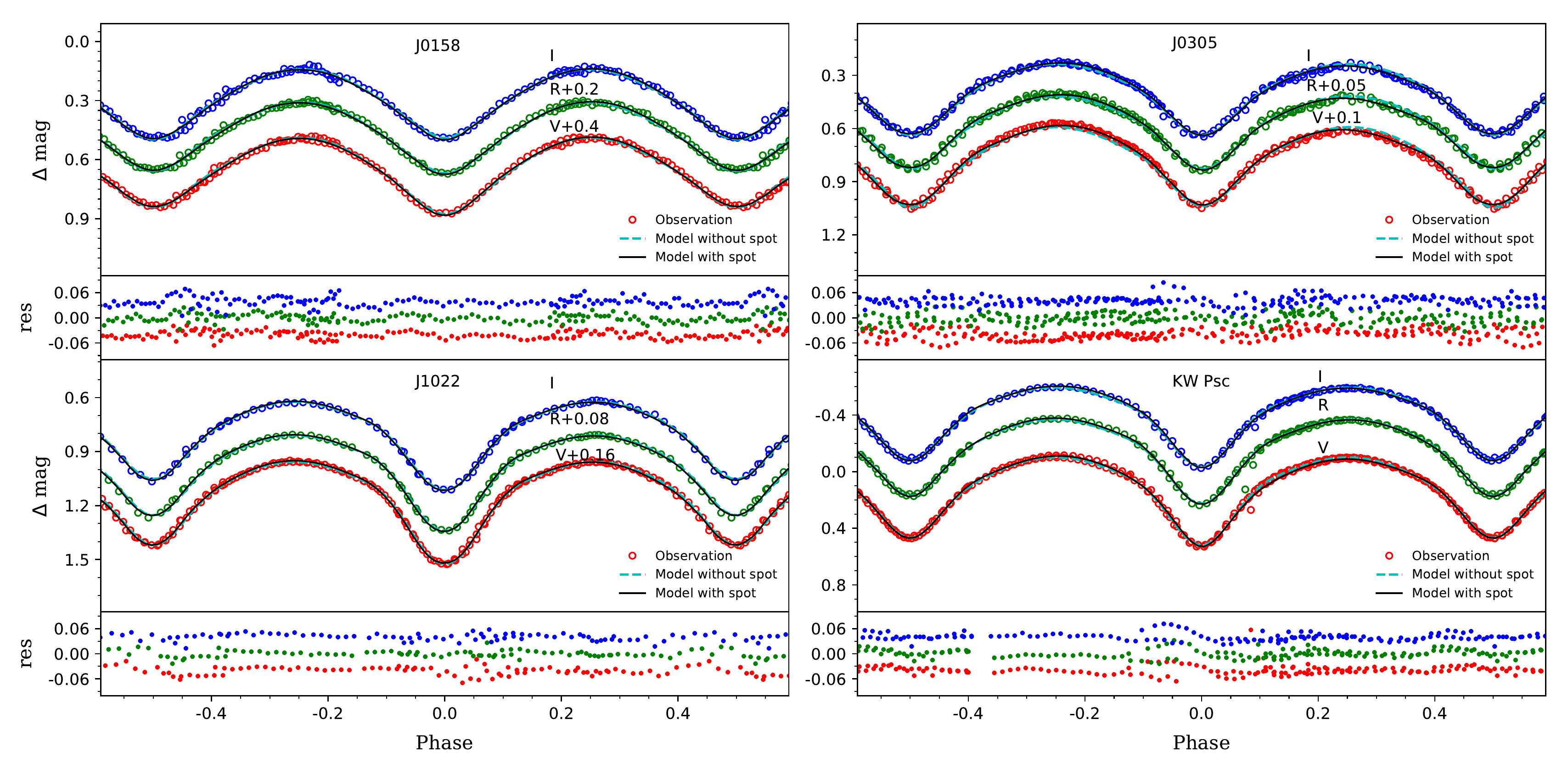}
\caption{The Observed and model fitted LCs in $VRI$ bands as shown by red, green and blue open circles. The lower panels of each plot show the residuals of the fitted model.}
\label{mo_fit}
\end{center}
\end{figure*}
For J0158, the estimated input $q$ is from 0.67($\pm$0.12) and primary $T_{eff}$ is 6156 ($\pm$35) K. The final photometric solutions show that the secondary $T_{eff}$ is less than the primary $T_{eff}$ by $\sim160$ K. We determined fill-out factors for primary and secondary components as 0.282 ($f_{1}$, $f_{2}$) respectively. The J0158 LCs show small asymmetry at phases 0.25 and 0.75. This is a well known effect of CBs and known as O'Connel effect \citep{1951PRCO....2...85O}. To understand this asymmetricity in the LCs of J0158, we considered a spot on primary while modeling it. It is not possible to identify the presence of spot on any component without Doppler imaging technique. Two different set of sport parameters can generate similar LC. The non-uniqueness of spot parameters obtained using photometric data alone is discussed previously by many authors. According to \cite{1999TJPh...23..357E}, the reasonable accuracy in spot parameters can be achieved only if photometric data accuracy is better than 0.0001 mag. We arbitrarily selected a cool spot for all the systems. The position and other spot parameters were decided on the basis of minimum cost function. The best fit model found that the spot was at co-latitude of $90^{o}$ and longitude of $145^{o}$. The position of spot was fixed while determining its radius and temperature ratio. The radius and $T_{spot}/T_{star}$ were estimated as $17^{o}$ and 0.93.

The observed LCs of J0305 have almost similar primary and secondary minima. The primary and secondary $T_{eff}$ are determined as 5125 ($\pm$41) and 5112 ($\pm$3) K, respectively. The temperature difference between components is $\sim$10 K which shows that they are in good thermal contact. J0305 also shows asymmetry in the observed LCs. The ($Max_{1}$-$Max_{2}$) for J0305 is about 0.04 mag. The fill-out factor for J0305 is 0.105. A cool spot on secondary was used while modeling the system J0305. Initially, spot was fixed at co-latitude of $90^{o}$ and longitude of $90^{o}$ but it was further moved towards the pole to get better fit. Finally, the best fit found the spot at co-latitude of $69^{o}$ and longitude of $75^{o}$. The radius and $T_{spot}/T_{star}$ are estimated as $23^{o}$ and 0.88. The rest of parameters are summarized in Table~\ref{mod_para}.
\begin{figure*}[!ht]
\begin{center}
\includegraphics[width=16.0cm, height=10.0cm]{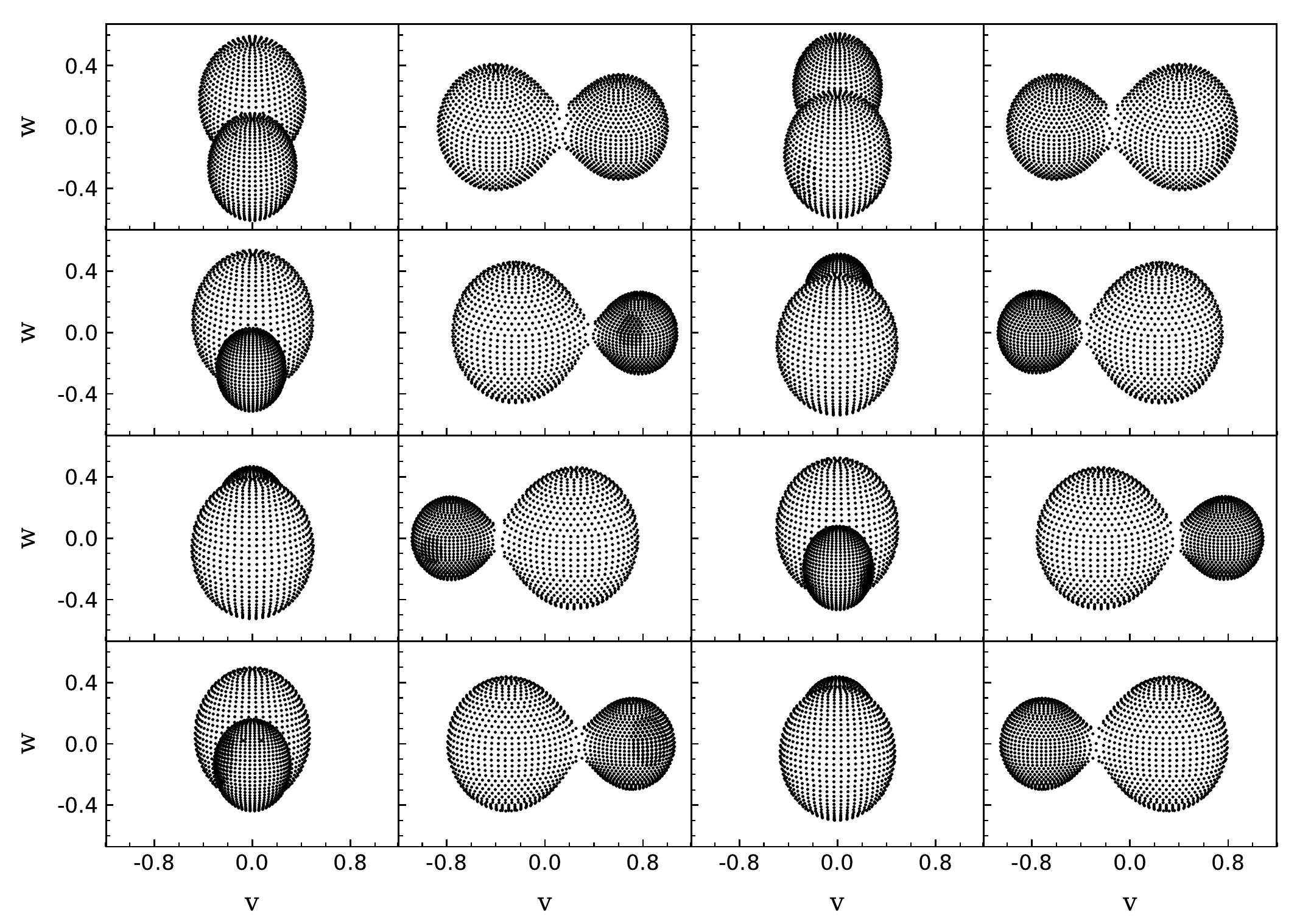}
\caption{Spot distribution on surface of eclipsing binaries. The first, second, third and fourth row from upper shows geometry of J0158, J0305, J1022 and KW Psc with spots at phases 0, 0.25, 0.50 and 0.75, respectively. }
\label{spots}
\end{center}
\end{figure*}
For J1022, it can be seen in Figure~\ref{lc_obs} that the primary and secondary minima are at different levels. The photometric solutions show that the secondary $T_{eff}$ is less than primary by $\sim300$ K. The fill-out factor was found to be 0.177 and 0.194 for both primary and secondary. Although LCs show very small asymmetricity, we still applied a cool spot on primary to improve our fit and determined the best fit parameters. The sum of squared residuals ($\Sigma~res^2$) reduced to 0.02 from 0.024 after including the spot. The radius and $T_{spot}/T_{star}$ are estimated as $19^{o}$ and 0.95. The spot position was found at co-latitude of $94^{o}$ and longitude of $235^{o}$ as shown in Figure~\ref{spots}.

For KW Psc, we obtained secondary $T_{eff}$ as 4830 ($\pm$3). It is about 90 K less than the primary $T_{eff}$, so, both the components are in good thermal contact. The fill-out factors were calculated as 0.192 ($f_{1}$) and 0.231 ($f_{2}$). The (Max$_{1}$-Max$_{2}$) for KW Psc is about -0.02. For this asymmetry, we used a cool spot on secondary at co-latitude of $76^{o}$ and longitude of $120^{o}$ with radius and $T_{spot}/T_{star}$ of $31^{o}$ and 0.96, respectively. The other parameters obtained from LCs fitting are given in Table~\ref{mod_para}. Figure~\ref{spots} illustrates the geometrical representation of the systems having spots on specific positions.
\input{tab07.tex}
\section{Physical Parameters}\label{phy_para}
The parameters like $q$, $i$, $f$, $L_{1}/(L_{1}+L_{2})$,  $r_{1}$,  $r_{2}$  are estimated by modeling of observed LCs. All the four sources are observed in GAIA. The GAIA parallaxes ($\pi$) given in Table~\ref{tar_info} are used to determine the absolute magnitude using :
\begin{equation}
M_{V}=m_{V}-5log(1000/\pi)+5-A_{V}
\end{equation}
where $M_{V}$, $m_{V}$, $\pi$ and $A_{V}$ represent absolute magnitude in V-band, apparent magnitude in V-band, parallax and extinction in V-band. The $\pi$ is in milli-arcsec. The $A_{V}$ is used from \cite{2011ApJ...737..103S} and average $m_{V}$ from \cite{2014yCat..22130009D}. The absolute V-band magnitudes are found to be 3.284, 5.581, 5.309 and 6.232 mag for J0158, J0305, J1022 and KW Psc, respectively. For estimating absolute bolometric magnitude ($M_{bol}$) from absolute magnitude, the bolometric corrections were used from \cite{2011yCat..21930001W} corresponding to the $T_{eff}$, metallicity and surface gravity of each system (-0.04, -0.24, -0.16 and -0.31 for J0158, J0305, J1022 and KW Psc, respectively). The $M_{bol}$ for J0158, J0305, J1022 and KW Psc are found to be 3.244, 5.341, 5.149 and 5.922 mag, respectively.
\begin{figure*}[!ht]
\begin{center}
\includegraphics[width=17cm, height=8cm]{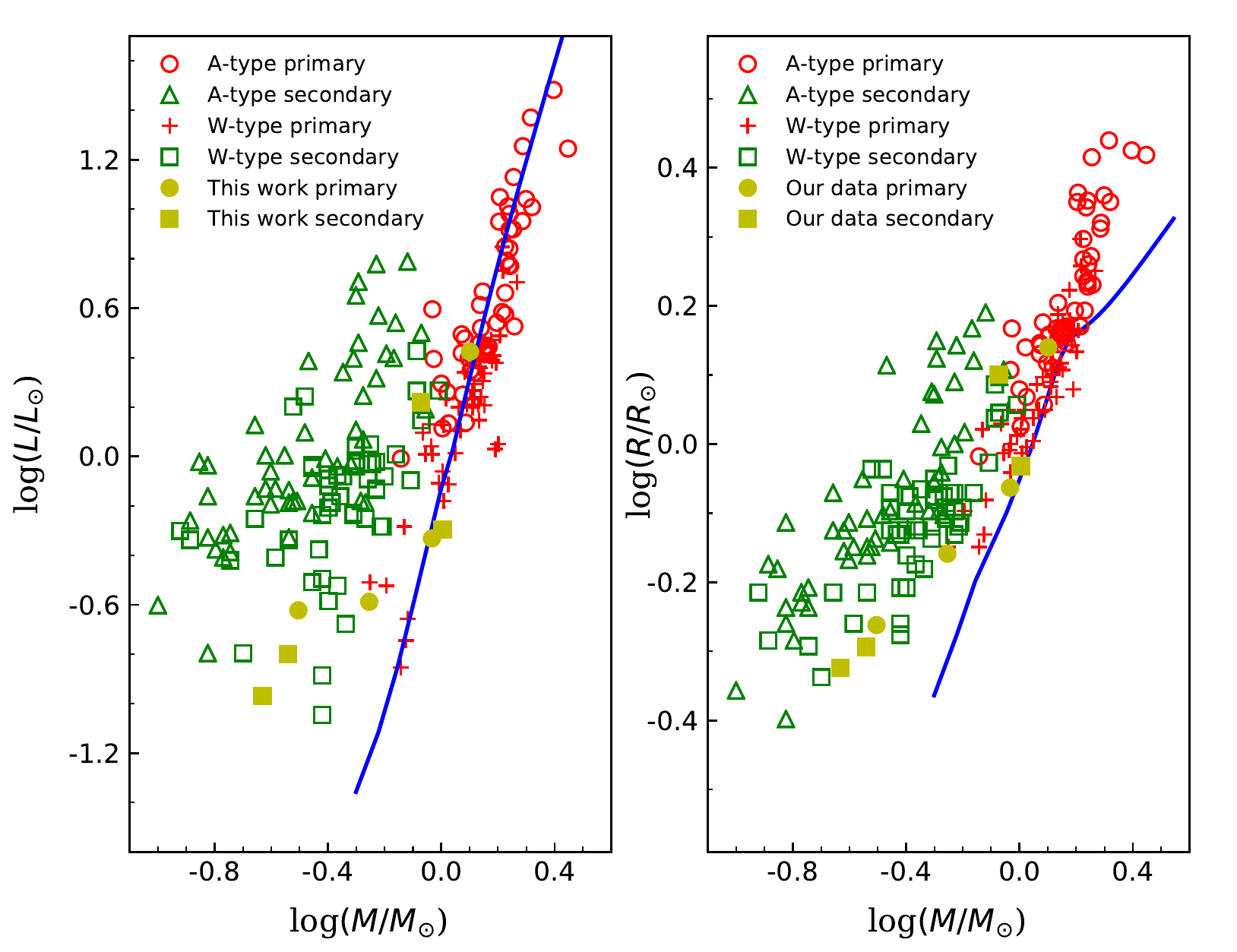}
\caption{The diagram shows the position of our systems with previously studies systems \citep{2013MNRAS.430.2029Y} on Mass-Luminosity and Mass- Radius planes. The continuous lines are ZAMS taken from \cite{2012yCat..35410041M} corresponding to z=0.014.}
\label{evol}
\end{center}
\end{figure*}

The total luminosity ($L_{1}+L_{2}$) was determined using :
\begin{equation}
\label{lu_eq}
L_{T}(L_{\odot}) = L_{1}+L_{2}=10^{-0.4(M_{bol}-M_{bol}{_\odot})}
\end{equation}
Here $M_{bol}{_\odot}$ is taken as 4.73 mag \citep{2010AJ....140.1158T}. Using the above equation, the total luminosity was calculated as 4.311, 0.625, 0.745 and 0.366 in $L_{\odot}$ units, for J0158, J0305, J1022 and KW Psc, respectively. The luminosity for individual component is determined by using the $L_{1}/(L_{1}+L_{2}$) obtained from LCs fitting in PHOEBE. For J0158, J0305, J1022 and KW Psc, $L_{1}$ is calculated as 2.655, 0.466, 0.238 and 0.258 $L_{\odot}$, respectively.

Total luminosity in terms of $T_{eff}$, relative radii of primary ($r_{1}$), relative radii of secondary ($r_{2}$) and separation of components (A) is given as:
\begin{equation}
\label{sm_eq}
L_{T}=T_{1}^{4}(Ar_{1})^{2}+T_{2}^{4}(Ar_{2})^{2}
\end{equation}
Here, $T_{1}$ and $T_{2}$ are in solar temperature units ($T_{\odot}$ = 5770 K). The separation, A, is in solar radius unit. The relative radii for primary or secondary is determined as:
\begin{equation}
\label{rel_ra}
r_{i}=(r_{pole} \times r_{side} \times r_{back})^{-1/3}
\end{equation}
Here $r_{pole}$, $r_{side}$, $r_{back}$ are obtained from the photometric LCs modeling. The $i$ is 1 for primary and 2 for secondary.

The $T_{1}$ and $L_{T}$ are already determined for all the systems. The $T_{2}$, $r_{1}$ and $r_{2}$ are determined by solution of LCs fittings. Finally, the Equation~\ref{sm_eq} was used for calculating the separation between components. The separation between two components in the sources J0158, J0305, J1022 and KW Psc are found to be 3.165, 1.772, 1.881 and 1.483 $R_{\odot}$, respectively.
\input{tab08.tex}

To determine the total mass ($M_{1}$+$M_{2}$) of the system, we used the Kepler's law. The constant factor in the Kepler's law is described in terms of $R_{\odot}$, day, $M_{\odot}$.
\begin{equation}
\dfrac{A^{3}}{P^{2}}=74.94(M_{1}+M_{2})
\end{equation} 
Here A, P, $M_{1}$ and $M_{2}$ are in units of $R_{\odot}$, days, $M_{\odot}$ and $M_{\odot}$, respectively. 

Using the earlier estimated values, we determined the $M_{1}+M_{2}$ as 2.108, 1.214, 1.325 and 0.790 $M_{\odot}$ for J0158, J0305, J1022 and KW Psc, respectively. The mass of individual components ($M_{1}$ and $M_{2}$) are determined using the $q$ value obtained through the LCs fitting. The radii of primary ($R_{1}$) and secondary ($R_{2}$) are determined from mean radii ($r_{i}$) and separation A by using following relation:
\begin{equation}
R_{i}=Ar_{i}
\end{equation}
The radii $R_{i}$ and A are in $R_{\odot}$ units. Using the $r_i$ estimated through the Equation (17), we calculated $R_i$ for each system. In Table~\ref{abs_para}, we give all the physical parameters determined for the four binary systems. The errors in these parameters are given in the parenthesis of each value which come as a result of error propagated through various equations used to determine the physical parameters considering the errors in individual parameters.

The position of these systems on $M$-$L$ and $M$-$R$ diagram  are shown in Figure~\ref{evol} along with the other previously studied EWs \citep[e.g.,][]{2013MNRAS.430.2029Y}. It can be seen that systems J0305, J1022 and KW Psc are near the group of W sub-type EWs. The primary component of all the systems is more closer to ZAMS as compared to the secondary. Using the mass and radius of previously studied cool contact binaries, \citet{1988MNRAS.231..341H} and \citet{1996A&A...311..523M} also noted similar trend in EWs. The secondary components are above ZAMS, which indicates that they have higher radius than a main sequence star with similary mass. As suggested by \cite{2004IAUS..219..967S}, this is not possible only due to energy transfer from primary to the secondary component, so, the secondary components must be more evolved with He-depletion cores.

\begin{figure*}[!ht]
\begin{center}
\subfigure{\includegraphics[width=8.5cm,height=5.8cm]{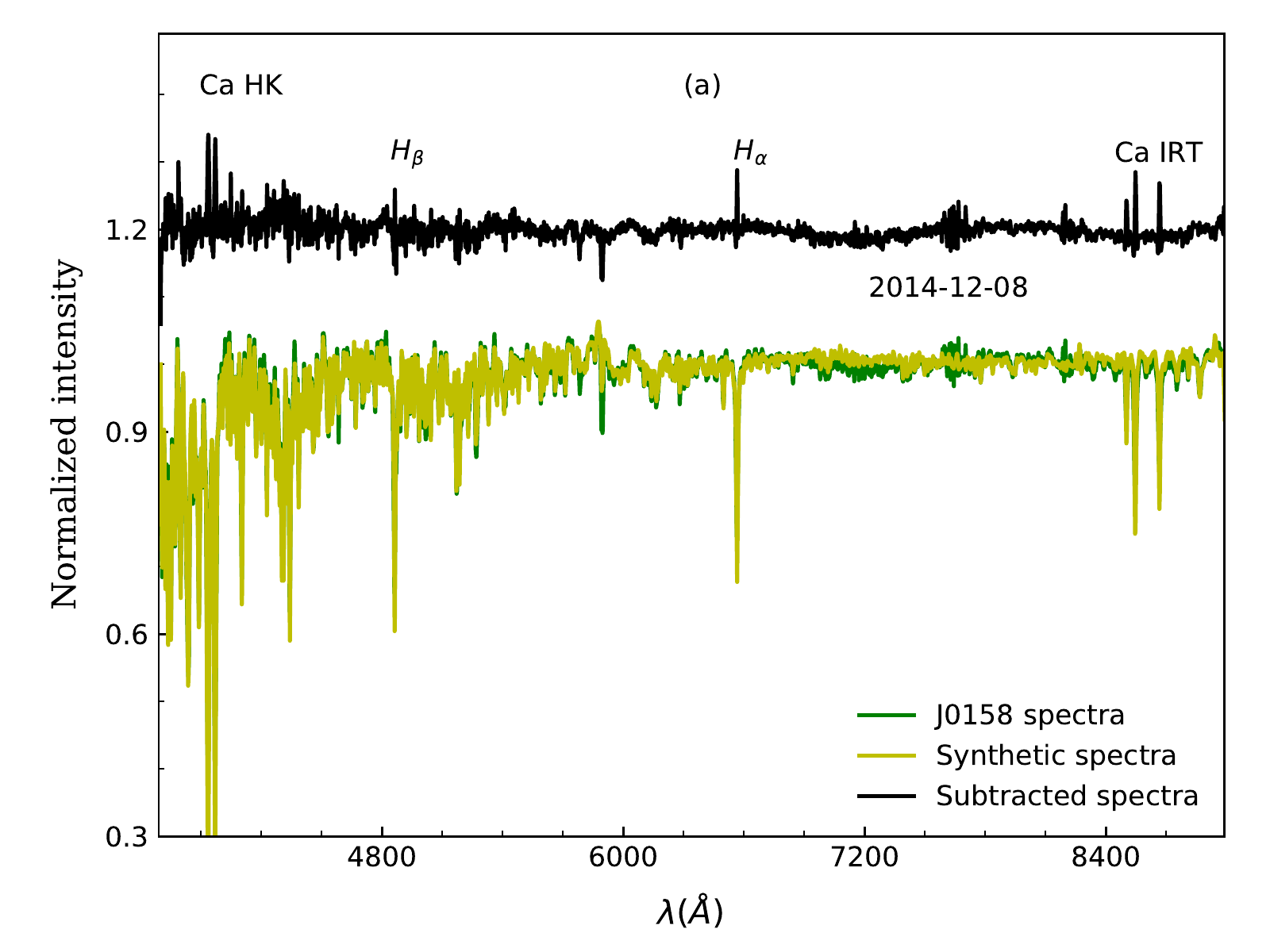}}
\subfigure{\includegraphics[width=8.5cm,height=5.8cm]{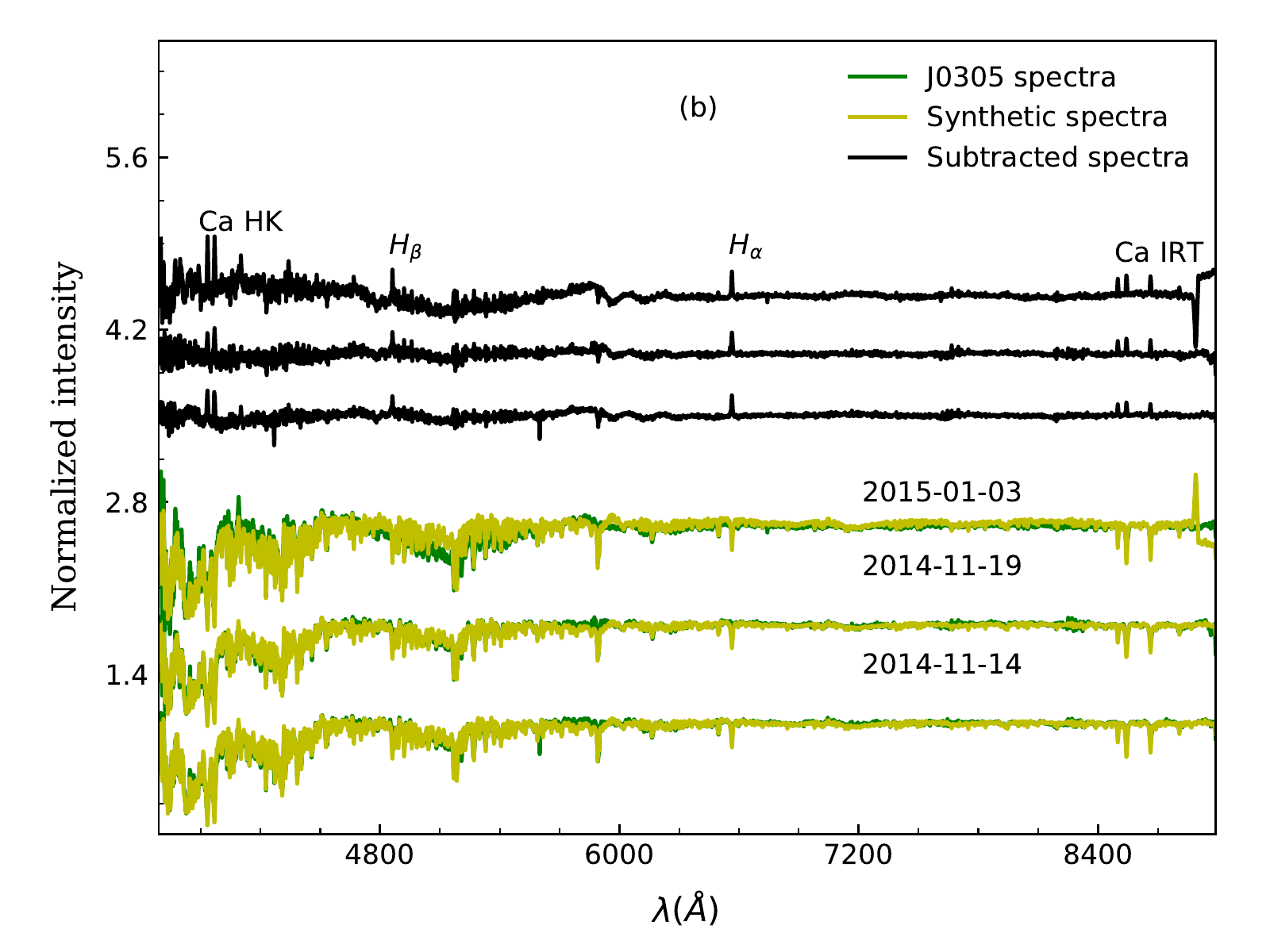}}
\caption{The LAMOST spectra for (a) J0158 and (b) J0305 over plotted with synthetic spectra. The black continuous line shows the subtracted spectra with excess emission lines.}
\end{center}
\label{spec1}
\end{figure*}
%
\begin{figure*}[!ht]
\begin{center}
\label{spec2}
\subfigure{\includegraphics[width=8.5cm,height=5.8cm]{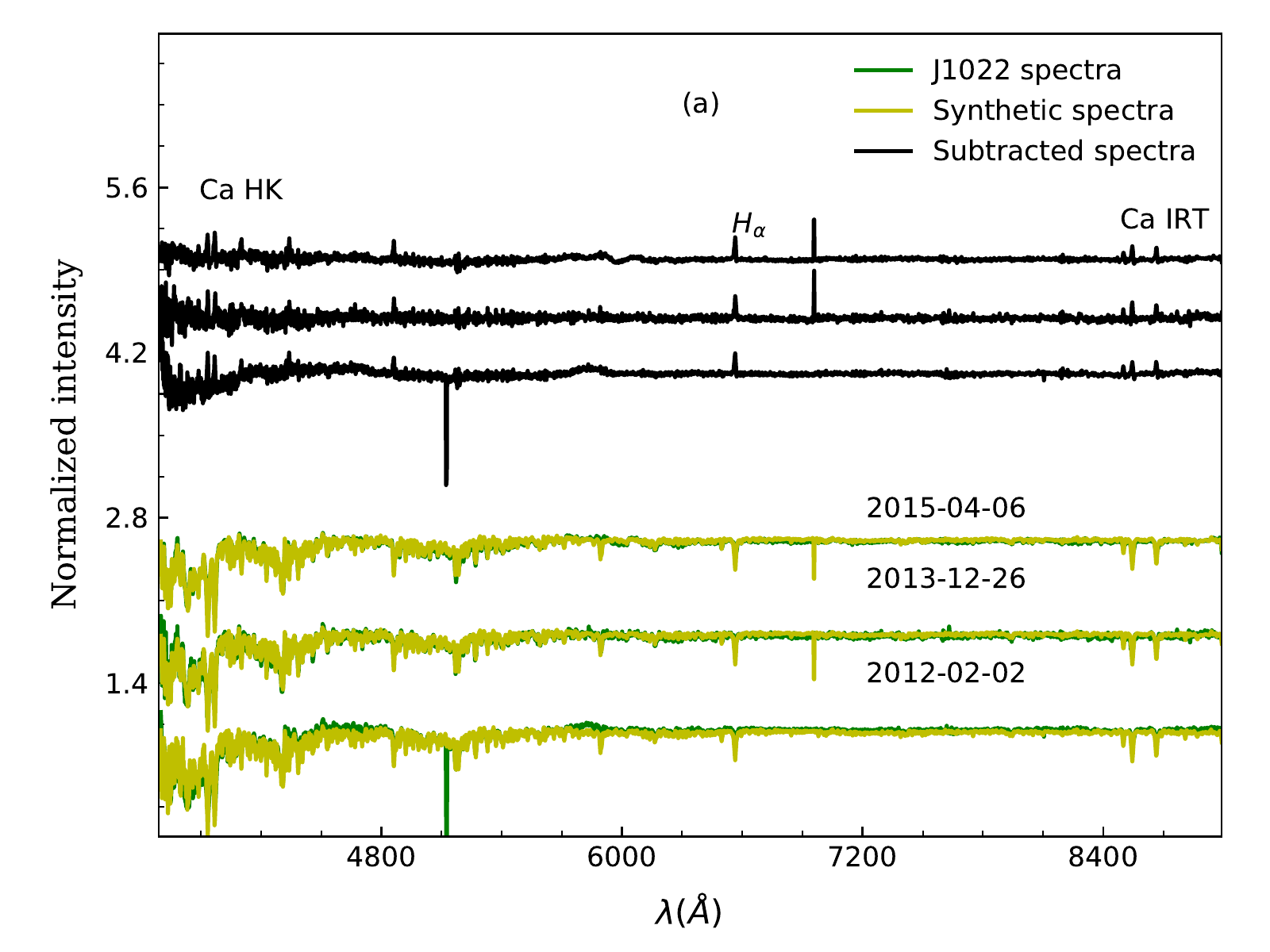}}
\subfigure{\includegraphics[width=8.5cm,height=5.8cm]{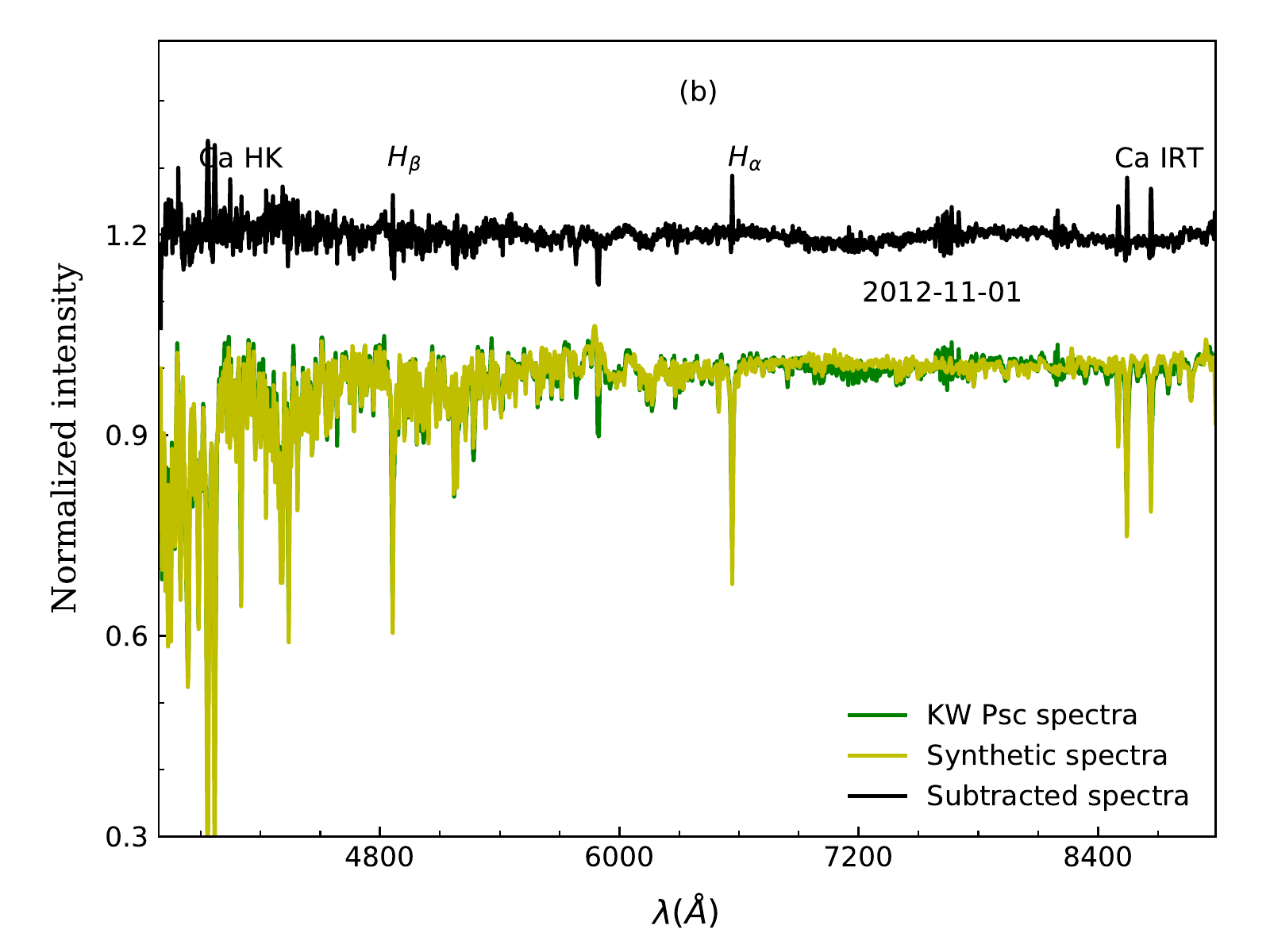}}
\caption{Same as in Figure~\ref{spec1} but for (a) J1022 and (b) KW Psc.}
\end{center}
\end{figure*}
%
\begin{figure*}[!ht]
\begin{center}
\includegraphics[width=15cm,height=8cm]{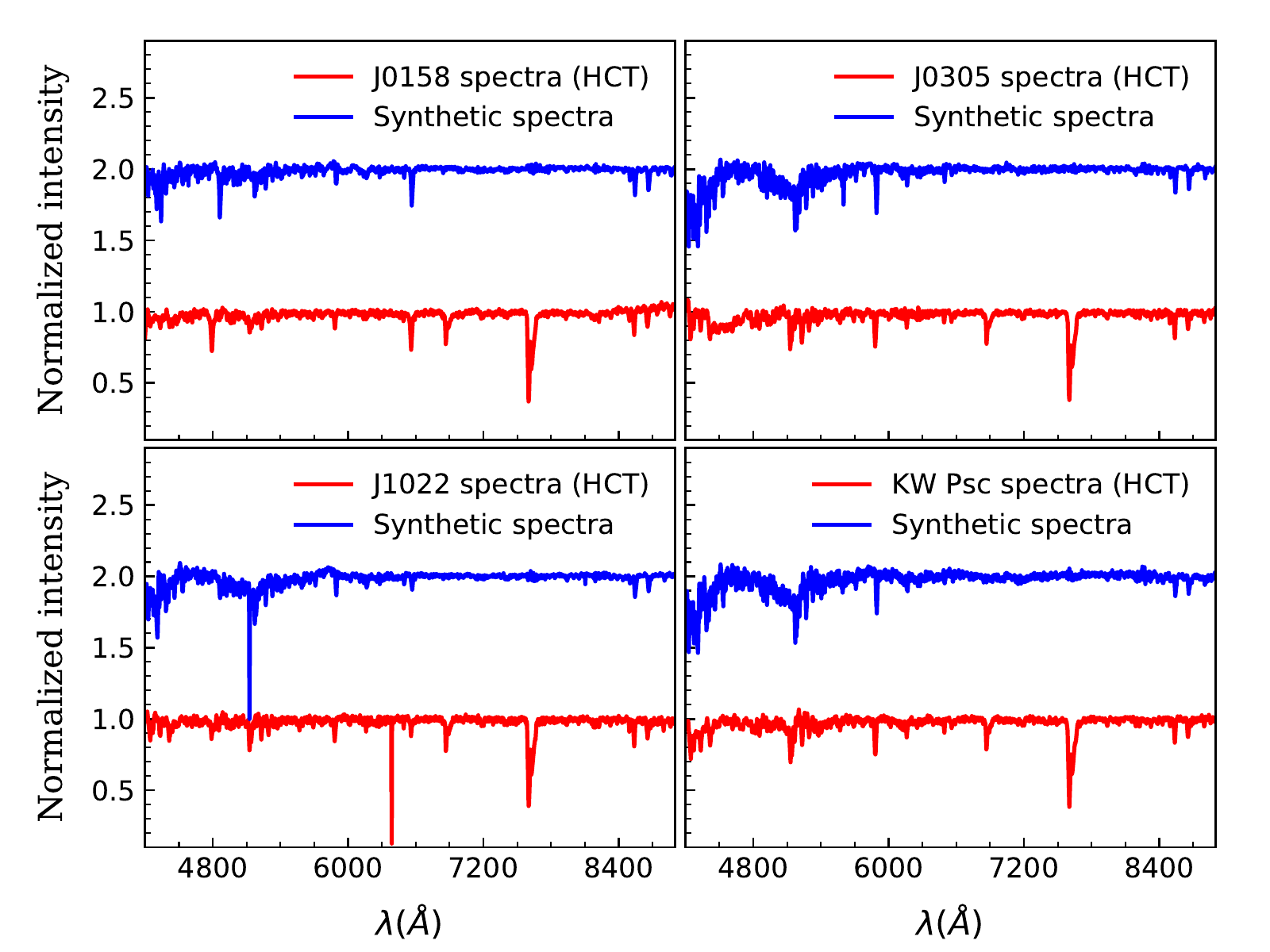}
\caption{The spectra of target sources obtained from the HCT (red color) and the synthesized spectra (blue color).}
\label{hct_sp}
\end{center}
\end{figure*}
\section{Chromospheric Activities}\label{ch_ac}
The phenomenon of magnetic activities is often seen in the late type rotating stars having convective envelopes which result in the formation of star spots, flares or plagues. The surface chromospheric activity depends on stellar rotation rate. The spectral emission lines $H_{\alpha}$, $H_{\beta}$, Mg $I~b$ triplet, Na $I~D_{1}~D_{2}$, $Ca~II~H~\&~K$, $Ca~II~IRT$ etc are optical and near-infrared indicators of chromospheric activity \citep{1984BAAS...16..893B, 1995A&AS..114..287M} and equivalent width of these lines provide a good measure of activity level in the late type rotating stars.

In case of binary stars, the total flux in spectra at a specific time contains contribution from chromospheric and photospheric flux of both the stars. The reconstruction of absorption profile and spectral subtraction techniques are commonly used for studying chromospheric activity of stars, however, former is widely used in the case of binary systems. The spectral subtraction technique is based on the assumption that the level of photospheric flux is almost same in the stars having similar spectral type. This suggests that an inactive star of similar spectral type can be used to estimate the photospheric flux contribution for an active star \citep{1984BAAS...16..893B, 1995A&AS..114..287M} which may be seen in the form of excess emission. 

The LAMOST spectra (for J0158, J0305 and KW Psc) and HCT spectra (for J0158, J0305, J1022 and KW Psc) are analysed for chromospheric activity signatures. The noise in the spectra can also produce some emission like features in the subtracted spectra, so, only the high SNR specra of objects and inactive stars are selected. The inactive stars having small rotational velocity are appropriate candidates for the template spectra due to less rotational broadening. Here, the synthetic spectra is constructed using the STARMOD program \citep{1984AJ.....89..549H, 1984BAAS...16..893B} which uses the inactive template spectra for both components of EWs and generate a composite spectra after introducing rotational broadening and radial velocity shifts. The stars HD 233641 \citep{2004yCat..21520261W}, HD 238130, HD 77712, HD 219829 \citep{2000yCat..41420275S} and BD+43 2328 \citep{2004ApJS..152..251V} are used for preparation of composite synthetic spectra. The spectra obtained after subtracting synthetic spectra from observed spectra are shown in Figures~\ref{spec1} to ~\ref{spec2}.
These spectra show emission in $H_{\alpha}$, $H_{\beta}$ and $Ca~II~H~\&~K$ and $Ca~II~IRT$ lines. Here, the spectra from HCT are shown in Figure~\ref{hct_sp} in red color while the synthetic spectra generated from LAMOST data are shown with blue color in the same figure. It is to be noted here that the spectral subtraction technique has not been applied on HCT spectra due to unavailability of comparison stars in those observations.

As we have earlier noticed asymmetry in the LCs of these systems which might have resulted due to some magnetic activities. The excess emission found in the differential spectra seems to confirm this notion. As SNR at both ends of spectra was poor we therefore could calculate the equivalent width of $H_{\alpha}$ line only. We used the spectral range from 653.5 nm to 659.5 nm and fitted Gaussian profile to determine the equivalent width. For J0158 and J0305 the equivalent widths were found to be $0.369\pm0.017$ and $1.031\pm0.018$. For J0305 and KW Psc we determined equivalent widths of $1.236\pm0.608$ and $1.206\pm0.042$. However, due to poor Gaussian fitting in other spectra, the equivalent widths could not be determined.
\section{Discussion and conclusion}\label{discu}
The detailed study of EWs are useful in understanding their formation mechanism and different evolution stages. A long-term photometric and spectroscopic study can thus throw light on their period change and associated processes like mass transfer, third body and spot evolution. In this study, we present the multi-band photometric and low resolution spectroscopic analysis of four EWs. Due to absence of radial velocity curves of these systems, mass ratios of the binary components are determined from photometric LCs with the $q$-search method. For all the systems $q$ is found to be less than 0.7 except J1022 for which a higher value of 3.23 is found. As the components are close to each other in EWs, the interaction between the components is quite common in these systems. This can result in a period change through mass transfer/loss among components. The presence of additional companion or long term cyclic magnetic activity is also prevalent for contact binaries which can cause cyclic variation in the (O-C) diagram. For the four EBs studied here, we could able to collect the TOM information during last 13-15 years only and with this limited time span it was very difficult to retrieve any specific information about the long term cyclic variations. However, a preliminary (O-C) analysis of these systems shows a change in period for three systems (J0158, J0305 and J1022) but no such variation was noticed in the case of KW Psc. The mass loss can be the reason for change in their periods hence we also calculated mass-transfer rates for these systems.

Asymmetry in LCs of all the systems is observed and level of asymmetry show change from $V$ to $I$ band with being maximum in $V$ band.The LCs from SuperWASP data show that these systems exhibit variation from positive O'Connell to negative O'Connell effect with passage of time. Even the depths of eclipsing minima are seen to be changing in these two systems. The analysis of present observations and the SuperWASP observations shows that for J0305 $Max_{1}-Max(2)$ varies from 0.06 to -0.06. Similarly, this difference for J1022 varies from 0.04 to -0.2. This behaviour indicates that spots are not fixed but form and move with time. 

Different empirical relations are available in literature for determining parameters like mass, radius, luminosity etc. However, these relations can be biased due to the specific EWs sample used during their formulation. Therefore, in the present analysis, we followed the procedure adopted by \cite{2020Ap&SS.365...71L} for calculating the physical parameters. Recently \cite{2020ApJS..247...50S} derived parameters of 2335 late-type contact binaries from the CSS survey including the system J0305. They found a mass-ratio of 0.19 for J0305 which is smaller than the estimated value of 0.31 in the present study. However, it should be noted that \citet{2020ApJS..247...50S} used only $V$ band data in their analysis and a primary component cooler by 150 K than the present value \textbf{hence some disagreement between the two estimates could be possible}. 
The total mass ($M_{t}$) determined for all the systems is above minimum total mass limit for EWs except for KW Psc. This indicates that significant amount of mass loss had taken place in KW Psc system in the past. As we have not found any period change in KW Psc during last 12 years, we believe that the observed low mass of this system can be due to any previous magnetic activities in the form of some burst. Nevertheless, a more detailed study is required to find the exact cause of total low mass in this system. As W UMa type systems with q$>$0.5 are W-subtype, so, J0158 can be classified as W-subtype but its spectral class and high temperature of primary suggests that it can be A-subtype system. Similarly, J0305 and KW Psc can be classified as W-subtypes on the basis of their spectral type but their q$<$0.5  and high temperature of primary places them into A-subtype category. The J1022 is found to be W-subtype EW. All the systems are shallow contact W UMa type.

The low resolution spectra from LAMOST and HCT for all the sources have been compared with the synthetic spectra using the spectral subtraction technique. The subtracted spectra for these systems show a small excess emission in $H_{\alpha}$, $H_{\beta}$ and Ca triplet region. Small emission is also visible in Ca HK region but considerable amount of noise in blue region makes it difficult to analyse. Although there is 3 to 4 year difference between LAMOST spectroscopic observations and our photometric observations, the presence of spots in LCs modeling can still be assumed an indirect proof of their activities. The equivalent widths of different lines in subtracted spectra can give measure of magnetic activity in these systems, however, further spectroscopic observations with better resolution at different phases will be more useful in the study of chromospheric activities in EBs.

\section{ACKNOWLEDGEMENTS}
The work presented here has been carried out under the DST project "INT/AUSTRIA/BMWF/P-14". We thank the staff of IAO, Hanle and CREST, Hosakote, that made these observations possible. The facilities at IAO and CREST are operated by the Indian Institute of Astrophysics, Bangalore. Guoshoujing Telescope (the Large Sky Area Multi-Object Fibre Spectroscopic Telescope LAMOST) is a National Major Scientific Project built by the Chinese Academy of Sciences. Funding for the project has been provided by the National Development and Reform Commission. LAMOST is operated and managed by the National Astronomical Observatories, Chinese Academy of Sciences. In this work we have also used the data from the European Space Agency (ESA) mission GAIA, processed by the GAIA Data Processing and Analysis Consortium (DPAC). This work also make use of the Two Micron All Sky Survey and SIMBAD database.

\bibliographystyle{yahapj}
\bibliography{Bibilography}

\end{document}

%% file: tab01.tex
\begin{table}[!ht]
\caption{Basic information about the sources taken from different surveys}   
\centering              
\label{tar_info}
\scriptsize         
\begin{tabular}{p{.32in}p{.35in}p{.35in}p{0.3in}p{0.2in}p{0.2in}p{0.2in}p{0.25in}}
\hline\hline     
Source& RA         &DEC        & Period   & V     &B-V    & J-K   & Parallax\\
      & (J2000)    &(J2000)    & (days)   & (mag) & (mag) & (mag) & (mas) \\
\hline
J0158 & 01:58:29.5 & +26:03:33 & 0.227665 & 12.71 & 0.629 & 0.305 & 1.445 \\
J0305 & 03:05:05.1 & +29:34:43 & 0.246984 & 12.18 & 0.952 & 0.668 & 6.344 \\
J1022 & 10:22:11.7 & +31:00:22 & 0.258468 & 12.53 & 0.792 & 0.494 & 3.702 \\
KW Psc& 22:58:31.7 & +05:52:23 & 0.234276 & 12.16 & 0.980 & 0.593 & 7.055 \\ 
\hline                  
\end{tabular}
\footnote{The (B-V) is taken from APASS survey \citep{2015AAS...22533616H}, (J-K) is taken from 2MASS survey \citep{2006AJ....131.1163S} and the parallax is from GAIA \citep{2020arXiv201201533G}.}
\end{table}

%% file: tab02.tex
\begin{table}
\caption{The observation log for the targets observed using 1.3-m DFOT.}
\centering
\label{log_phot}
\scriptsize
\begin{tabular}{p{.2in}p{.47in}p{.4in}p{0.4in}p{0.38in}p{0.52in}p{0.1in}}
\hline
Object&  Date of & Start Jd   &   End Jd    & Total  & Exposure       & Obs.      \\
      &obs.  &            &             & frames & time           & time      \\
      &          & (2450000+) & (2450000+)  &(V,R,I)& (sec)           &   (hrs)     \\
\hline
      &2018-11-20& 8443.1216  &  8443.3557  & 51, 50, 50   & 120,50-60,50   &    5.62     \\
J0158 &2018-12-27& 8480.1687  &  8480.2068  & 06, 06, 06   & 180,120,80     &    0.91     \\
      &2019-10-14& 8771.1721  &  8771.4613  & 84, 84, 84   & 40, 25, 20     &    6.94     \\
\hline
      &2018-11-26& 8449.1484  &  8449.3725  & 57, 56, 55   & 75, 35, 30     &    5.38     \\
      &2018-12-01& 8454.3645  &  8454.3779  & 04, 04, 04   & 75, 35, 30     &    0.32     \\
      &2018-12-21& 8474.2458  &  8474.3154  & 18, 18, 18   & 75, 35, 30     &    1.67     \\
J0305 &2018-12-22& 8475.0528  &  8475.1139  & 15, 15, 15   & 75, 35, 30     &    1.47     \\
      &2018-12-27& 8480.2023  &  8480.2422  & 10, 10, 10   & 75, 35, 30     &    0.96     \\
      &2019-01-17& 8501.0352  &  8501.1620  & 32, 32, 32   & 75, 35, 30     &    3.04     \\
      &2019-11-10& 8798.1963  &  8798.4309  & 66, 66, 66   & 75, 35, 30     &    5.63     \\
\hline
      &2019-03-19& 8562.1683  &  8562.2574  & 21, 20, 20   & 60, 35, 30     &    2.14     \\
      &2019-03-20& 8563.1543  &  8563.1805  & 08, 07, 07   & 60, 35, 30     &    0.63     \\
J1022 &2019-03-21& 8564.1612  &  8564.2565  & 26, 27, 27   & 60, 35, 30     &    2.29     \\
      &2019-04-01& 8575.1951  &  8575.2849  & 24, 24, 24   & 60, 40, 30     &    2.16     \\
      &2019-04-02& 8576.2056  &  8576.2341  & 08, 08, 08   & 60, 40, 30     &    0.68     \\
\hline
      &2018-10-11& 8403.0627  &  8403.1523  & 25, 25, 25   & 30, 25, 20     &    2.15     \\
      &2018-10-12& 8404.1082  &  8404.1592  & 15, 15, 15   & 30, 25, 20     &    1.22     \\
      &2018-10-20& 8412.0703  &  8412.2057  & 42, 41, 41   & 30, 25, 20     &    3.25     \\
KW    &2018-11-26& 8449.0536  &  8449.1280  & 20, 20, 20   & 70, 25, 20     &    1.79     \\
Psc   &2018-12-01& 8454.0309  &  8454.0706  & 10, 10, 10   & 70, 25, 20     &    0.95     \\
      &2018-12-27& 8480.0801  &  8480.1280  & 15, 15, 15   & 30, 25, 20     &    1.50     \\
      &2019-10-14& 8771.1260  &  8771.1495  & 07, 07, 07   & 30, 30, 30     &    0.55     \\
\hline                                                    
 \end{tabular}       
 \end{table}

%% file: tab03.tex
\begin{table}[!ht]
\caption{Parameters of targets from the LAMOST data}             
\label{tar_lamost}
\centering   
\scriptsize       
\begin{tabular}{p{.39in}p{.49in}p{.2in}p{0.28in}p{0.25in}p{0.27in}p{0.25in}}
\hline   
Targets & Date        & $T_{eff}$& Sub & logg  & Fe/H   & SNR\\
        &             &      (K)   &class     &       & (dex)  & \\
\hline
J0158   & 08-12-2014 & 6151       & F7      & 4.033 &  0.178 & 428.44\\ 
\hline
        & 14-11-2014 & 4917       & G9      & 4.451 & -0.423 & 239.10\\ 
J0305   & 19-11-2014 & 4839       & G9      & 4.358 & -0.406 & 189.31\\ 
        & 03-01-2015 & 4721       & K5      & 4.410 & -0.490 & 276.60\\ 
\hline
        & 02-02-2012 & 5382       & G8      & 4.317 & -0.259 & 245.76\\ 
J1022   & 26-12-2013 & 5211       & G8      & 4.142 & -0.380 & 58.97\\ 
        & 06-04-2015 & 5305       & G7      & 4.306 & -0.375 & 232.12\\ 
\hline
KW Psc  & 01-11-2012 & 4822       & G9      & 4.485 & -0.414 & 63.41\\ 
\hline                  
\end{tabular}
\end{table}

%% file: tab04.tex
\begin{table}
\caption{The log of spectroscopic observations for the targets observed using 2-m HCT.}
\label{log_spec}
\scriptsize
\centering
\begin{tabular}{ccccccc}
\hline
Object &   Date     &   Mid-UT   &   Mid-UT     & Exposure  & SNR   \\
       &            & (for GR7)  & (for GR8)    & (Sec)     &       \\
\hline
J0158  & 2019-11-17 & 14:02      &  14:33       & 1500      & 106   \\
J0305  & 2019-11-17 & 15:07      &  15:33       & 1500      & 92    \\
J1022  & 2019-11-17 & 22:31      &  22:56       & 1500      & 111   \\
KW Psc & 2019-11-17 & 12:57      &  13:23       & 1500      & 100  \\
\hline                                                    
 \end{tabular}       
 \end{table}

%% file: tab05.tex
\begin{table}[!ht]
\caption{TOMs estimated for J0158, J0305, J1022 and KW Psc using data from different surveys}
\label{OC_info}
\centering
\scriptsize
\begin{tabular}{p{.33in}p{.38in}p{.38in}p{0.1in}p{0.3in}p{0.35in}p{0.35in}p{0.1in}}
\hline
ID     & $HJD_{o}$  & Error   &Min& Cycle & $(O-C)_{1}$ & $(O-C)_{2}$ & Ref \\
       &(2450000+)  &         &   &       &  (days)     & (days)      &     \\
\hline
J0158  & 3229.67908 & 0.00042 & p &  0   &-0.00571 &-0.00679 & 1 \\
J0158  & 3232.64338 & 0.00038 & s &  6.5 &-0.00108 &-0.00215 & 1 \\
...    & ...        & ...     &...&  ... & ...     & ...     & ..\\
J0305  & 3228.68060 & 0.00041 & p &-3469 & 0.02795 & 0.00041 & 1 \\
J0305  & 3229.66817 & 0.00048 & p &-3465 & 0.02759 & 0.00005 & 1 \\
...    & ...        & ...     &...&  ... & ...     & ...     & ..\\
J1022  & 4075.64602 & 0.00022 & p &-17365&-0.01739 & 0.00816 & 1 \\
J1022  & 4100.71649 & 0.00015 & p &-17268&-0.01984 & 0.00333 & 1 \\
...    & ...        & ...     &...&  ... & ...     & ...     & ..\\
KW Psc & 4354.42230 & 0.00010 & p & -2819& 0.00596 & 0.00095 & 2 \\
KW Psc & 5014.84300 & 0.00000 & p & 0    &-0.00225 &-0.00030 & 3 \\
...    & ...        & ...     &...&  ... & ...     & ...     & ..\\

\hline                  
\end{tabular}

\vspace{1ex}
{\raggedright Here [1], [2] and [3] show TOMs estimated by SuperWASP, \cite{2010IBVS.5922....1G} and \cite{2010IBVS.5920....1D}. This is only sample table.\par}
\end{table}

%% file: tab06.tex
\begin{table}[!ht]
\caption{$T_{eff}$ (in $^{\circ}C$) determined from different empirical relations and LAMOST data.}             
\label{all_temp}      
\centering
\scriptsize
\begin{tabular}{c c c c c c}    
\hline\hline     
J0158 & J0305 & J1022 & KW Psc & Ref \\
\hline
 6140$\pm$105  & 4829$\pm$105 & 5440$\pm$118 & 5047$\pm$138 & 1   \\
 6274$\pm$11   & 5445$\pm$51  & 5380$\pm$3   & 4822$\pm$6   & 2   \\
 6061          & 5399         & 5356         & 4995         & 3   \\
 6151$\pm$11   & 4826$\pm$38  & 5299$\pm$69  & 4822$\pm$69  & 4   \\
 6156$\pm$35   & 5125$\pm$41  & 5369$\pm$46  & 4921$\pm$51  & 5     \\

\hline                  
\end{tabular}
\vspace{1ex}

{\raggedright Here [1], [2], [3], [4] and [5] represents the $T_{eff}$ determined using relation by \cite{2007MNRAS.380.1230C}, \cite{1994ApJ...434..277W}, \cite{2010AJ....140.1158T}, LAMOST DR5 and the mean temperature used in present analysis. \par}
\end{table}

%% file: tab07.tex
\begin{table}[!ht]
\caption{Results from LC fitting for all the four systems. The values in parentheses are errors in last digits.}             
\label{mod_para}      
\centering
\scriptsize
\begin{tabular}{l c c c c}    
\hline\hline     
Parameters      & J0158     & J0305 & J1022 & KW Psc  \\
\hline
q ($M_{2}/M_{1}$)        & 0.67(12)   & 0.31(1)    & 3.23(14)   & 0.42(1)    \\
i ($^{\circ}$)           & 64.58(44)  & 71.65(18)  & 75.41(72)  & 78.74(17)  \\
$T_{1}$ (K)              & 6156(35)   & 5125(41)   & 5369(46)   & 4921(51)   \\
$T_{2}$ (K)              & 5991(25)   & 5112(3)    & 5083(6)    & 4830(3)    \\
$\Omega_{1}$             & 3.08(21)   & 2.47(1)    & 6.81(34)   & 2.67(2)    \\
$\Omega_{2}$             &$\Omega_{1}$&$\Omega_{1}$& 6.80(2)    & 2.66(1)    \\
$\Omega_{in}$            & 3.19       & 2.49       & 6.92       & 2.72       \\
$\Omega_{out}$           & 2.80       & 2.30       & 6.30       & 2.46       \\
$L_{1}/(L_{1}+L_{2})$(V) & 0.616      & 0.746      & 0.320      & 0.706      \\
$L_{1}/(L_{1}+L_{2})$(R) & 0.612      & 0.745      & 0.308      & 0.701      \\
$L_{1}/(L_{1}+L_{2})$(I) & 0.608      & 0.745      & 0.301      & 0.698      \\
$r_{1}$ (pole)           & 0.407      & 0.457      & 0.271      & 0.438      \\
$r_{1}$ (side)           & 0.433      & 0.492      & 0.283      & 0.469      \\
$r_{1}$ (back)           & 0.471      & 0.518      & 0.322      & 0.499      \\
$r_{1}$ (average)        & 0.436      & 0.488      & 0.291      & 0.468      \\
$r_{2}$ (pole)           & 0.341      & 0.268      & 0.461      & 0.298      \\
$r_{2}$ (side)           & 0.459      & 0.279      & 0.497      & 0.312      \\
$r_{2}$ (back)           & 0.403      & 0.315      & 0.525      & 0.353      \\
$r_{2}$ (average)        & 0.398      & 0.287      & 0.494      & 0.320      \\
$f_{1}$,$f_{2}$          & 0.282      & 0.105      &0.177,0.194 & 0.192,0.231\\
\hline                  
\end{tabular}
\vspace{1ex}
\end{table}

%% file: tab08.tex
\begin{table}[!ht]
\caption{Absolute parameters determined using GAIA parallax and LC solutions for all the four systems. The values in parentheses are errors in last digits.}             
\label{abs_para}
\scriptsize
\centering          
\begin{tabular}{l c c c c}    
\hline\hline     
Parameters      & J0158     & J0305 & J1022 & KW Psc  \\
\hline
a($R_{\odot}$)           & 3.165(121) & 1.772(54) & 1.881(49) & 1.483(47)  \\
$M_{1}$ ($M_{\odot}$)    & 1.262(171) & 0.927(85) & 0.313(27) & 0.557(53)  \\
$M_{2}$ ($M_{\odot}$)    & 0.846(189) & 0.287(28) & 1.011(97) & 0.234(23)  \\
$R_{1}$ ($R_{\odot}$)    & 1.381(53)  & 0.865(26) & 0.547(14) & 0.694(22)  \\
$R_{2}$ ($R_{\odot}$)    & 1.259(48)  & 0.509(16) & 0.929(24) & 0.474(15)   \\
$L_{1}$ ($L_{\odot}$)    & 2.655(182) & 0.466(17) & 0.238(9)  & 0.258(2)   \\
$L_{2}$ ($L_{\odot}$)    & 1.655(113) & 0.159(6)  & 0.507(19) & 0.107(1)   \\
\hline                  
\end{tabular}
\vspace{1ex}

\end{table}